\documentstyle[12pt,epsf]{article}

\def\Preprint{\vspace*{-1.5cm}  
   \mbox{}      \hfill 
  FTUV/97-22 \\ \mbox{}\hfill 
  IFIC/97-22 \\  }
\catcode`\@=11
\long\def\@makefntext#1{
\protect\noindent \hbox to 3.2pt {\hskip-.9pt  
$^{{\ninerm\@thefnmark}}$\hfil}#1\hfill}		

\def\@makefnmark{\hbox to 0pt{$^{\@thefnmark}$\hss}}  
	
\def\ps@myheadings{\let\@mkboth\@gobbletwo
\def\@oddhead{\hbox{}
\rightmark\hfil\ninerm\thepage}   
\def\@oddfoot{}\def\@evenhead{\ninerm\thepage\hfil
\leftmark\hbox{}}\def\@evenfoot{}
\def\sectionmark##1{}\def\subsectionmark##1{}}

\renewcommand{\thefootnote}{\fnsymbol{footnote}}

\newcounter{sectionc}\newcounter{subsectionc}\newcounter{subsubsectionc}
\renewcommand{\section}[1] {\vspace*{0.6cm}  
        \refstepcounter{sectionc}
\setcounter{subsectionc}{0}\setcounter{subsubsectionc}{0}\noindent 
	{\normalsize\bf\thesectionc. #1}\par\vspace*{0.4cm}}
\renewcommand{\subsection}[1] {\vspace*{0.6cm}\addtocounter{subsectionc}{1} 
	\setcounter{subsubsectionc}{0}\noindent 
	{\normalsize\it\thesectionc.\thesubsectionc. #1}\par\vspace*{0.4cm}}
\renewcommand{\subsubsection}[1]
{\vspace*{0.6cm}\addtocounter{subsubsectionc}{1}
	\noindent {\normalsize\rm\thesectionc.\thesubsectionc.\thesubsubsectionc. 
	#1}\par\vspace*{0.4cm}}
\newcommand{\nonumsection}[1] {\vspace*{0.6cm}\noindent{\normalsize\bf #1}
	\par\vspace*{0.4cm}}
					         
\newcounter{appendixc}
\newcounter{subappendixc}[appendixc]
\newcounter{subsubappendixc}[subappendixc]

\renewcommand{\appendix}[1] {\vspace*{0.6cm}
        \refstepcounter{appendixc}
        \setcounter{figure}{0}
        \setcounter{table}{0}
        \setcounter{equation}{0}
        \renewcommand{\thefigure}{\Alph{appendixc}.\arabic{figure}}
        \renewcommand{\thetable}{\Alph{appendixc}.\arabic{table}}
        \renewcommand{\theappendixc}{\Alph{appendixc}}
        \renewcommand{\theequation}{\Alph{appendixc}.\arabic{equation}}
        \noindent{\bf Appendix \theappendixc #1}\par\vspace*{0.4cm}}

\def\abstracts#1{{
	\centering{\begin{minipage}{13truecm}     
     \footnotesize\baselineskip=12pt\noindent
	\centerline{\footnotesize ABSTRACT}\vspace*{0.3cm}
	\parindent=0pt #1
	\end{minipage}}\par}} 

\renewenvironment{thebibliography}[1]
	{\begin{list}{\arabic{enumi}.}
	{\usecounter{enumi}\setlength{\parsep}{0pt}
\setlength{\leftmargin 1.25cm}{\rightmargin 0pt}
	 \setlength{\itemsep}{0pt} \settowidth
	{\labelwidth}{#1.}\sloppy}}{\end{list}}

\newcounter{itemlistc}
\newcounter{romanlistc}
\newcounter{alphlistc}
\newcounter{arabiclistc}

\newcommand{\fcaption}[1]{
        \refstepcounter{figure}
        \setbox\@tempboxa = \hbox{\footnotesize Fig.~\thefigure. #1}
        \ifdim \wd\@tempboxa > 6in
           {\begin{center}
        \parbox{6in}{\footnotesize\baselineskip=12pt Fig.~\thefigure. #1}
            \end{center}}
        \else
             {\begin{center}
             {\footnotesize Fig.~\thefigure. #1}
              \end{center}}
        \fi}

\newcommand{\tcaption}[1]{
        \refstepcounter{table}
        \setbox\@tempboxa = \hbox{\footnotesize Table~\thetable. #1}
        \ifdim \wd\@tempboxa > 6in
           {\begin{center}
        \parbox{6in}{\footnotesize\baselineskip=12pt Table~\thetable. #1}
            \end{center}}
        \else
             {\begin{center}
             {\footnotesize Table~\thetable. #1}
              \end{center}}
        \fi}

\def\@cite#1#2{\unskip\nobreak\relax
    \def\@tempa{$\m@th^{\hbox{\the\scriptfont0 #1}}$}%
    \futurelet\@tempc\@citexx}
\def\@citexx{\ifx.\@tempc\let\@tempd=\@citepunct\else
    \ifx,\@tempc\let\@tempd=\@citepunct\else
    \let\@tempd=\@tempa\fi\fi\@tempd}
\def\@citepunct{\@tempc\edef\@sf{\spacefactor=\the\spacefactor\relax}\@tempa
    \@sf\@gobble}

\def\citenum#1{{\def\@cite##1##2{##1}\cite{#1}}}
\def\citea#1{\@cite{#1}{}}

\newcount\@tempcntc
\def\@citex[#1]#2{\if@filesw\immediate\write\@auxout{\string\citation{#2}}\fi
  \@tempcnta\z@\@tempcntb\m@ne\def\@citea{}\@cite{\@for\@citeb:=#2\do
    {\@ifundefined
       {b@\@citeb}{\@citeo\@tempcntb\m@ne\@citea\def\@citea{,}{\bf ?}\@warning
       {Citation `\@citeb' on page \thepage \space undefined}}%
    {\setbox\z@\hbox{\global\@tempcntc0\csname b@\@citeb\endcsname\relax}%
     \ifnum\@tempcntc=\z@ \@citeo\@tempcntb\m@ne
       \@citea\def\@citea{,}\hbox{\csname b@\@citeb\endcsname}%
     \else
      \advance\@tempcntb\@ne
      \ifnum\@tempcntb=\@tempcntc
      \else\advance\@tempcntb\m@ne\@citeo
      \@tempcnta\@tempcntc\@tempcntb\@tempcntc\fi\fi}}\@citeo}{#1}}
\def\@citeo{\ifnum\@tempcnta>\@tempcntb\else\@citea\def\@citea{,}%
  \ifnum\@tempcnta=\@tempcntb\the\@tempcnta\else
   {\advance\@tempcnta\@ne\ifnum\@tempcnta=\@tempcntb \else \def\@citea{--}\fi
    \advance\@tempcnta\m@ne\the\@tempcnta\@citea\the\@tempcntb}\fi\fi}

 1
 1
 1

\font\ninerm=cmr9

\renewcommand{\theequation}{\arabic{sectionc}.\arabic{equation}}
\def\refjl#1#2#3#4#5#6{\bibitem{#1} #2, {\it #3} {\bf #4} (#5) #6.}
\def\refbk#1#2#3#4{\bibitem{#1} #2, {\it #3}, #4.}
\def\etal{{\it et al}}
\def\NP{Nucl. Phys.}
\def\NPPS{Nucl. Phys. B (Proc. Suppl.)}
\def\PL{Phys. Lett.}
\def\PRL{Phys. Rev. Lett.}
\def\PR{Phys. Rev.}
\def\PRep{Phys. Rep.}
\def\ZP{Z. Phys.}
\def\MPL{Mod. Phys. Lett.}
\def\MP{Int. J. Mod. Phys.}
\def\JPG{J. Phys. G: Nucl. Phys.}           

\def\APNY{Ann. Phys., NY}

\def\RPP{Rep. Prog. Phys.}
\def\ARNPS{Ann. Rev. Nucl. Part. Sci.}
\def\PTP{Prog. Theor. Phys.}
\def\PPNP{Prog. Part. Nucl. Phys.}
\def\CPC{Comput. Phys. Commun.}
\newcommand{\eqn}[1]{(\ref{#1})}
\newcommand{\be}{\begin{equation}}
\newcommand{\ee}{\end{equation}}
\newcommand{\no}{\nonumber}
\newcommand{\bel}[1]{\be\label{#1}}
\newcommand{\ba}{\begin{array}{c}}
\newcommand{\bat}{\begin{array}{cc}}
\newcommand{\ea}{\end{array}}
\newcommand{\beqn}{\begin{eqnarray}}
\newcommand{\eeqn}{\end{eqnarray}}

\newcommand{\bi}{\begin{itemize}}
\newcommand{\ei}{\end{itemize}}

\newcommand{\rms}{\rm\scriptsize}

\newcommand{\lsim}{~{}_{\textstyle\sim}^{\textstyle <}~}
\newcommand{\lrder}{\stackrel{\leftrightarrow}{\partial}}

\newcommand{\cL}{{\cal L}}

\newcommand{\cO}{{\cal O}}
\newcommand{\cP}{{\cal P}}

\newcommand{\cA}{{\cal A}}

\begin{document}
\Preprint

\centerline{\large\bf TAU PHYSICS\footnote{\protect\bf
To appear in ``Heavy Flavours II'', eds. A.J. Buras and M. Lindner
(World Scientific, 1997).
}}
\baselineskip=22pt
\centerline{\footnotesize A. PICH}
\baselineskip=13pt
\centerline{\footnotesize\it  Departament de F\'{\i}sica Te\`orica,
IFIC,  CSIC -- Universitat de Val\`encia}
\baselineskip=12pt
\centerline{\footnotesize\it  Dr. Moliner 50, E-46100 Burjassot,
Val\`encia, Spain}
\vspace*{0.3cm}
\vspace*{0.7cm}
\abstracts{The pure leptonic or semileptonic character of $\tau$
decays makes  them a good laboratory to test the structure of the
weak currents  and the universality of their couplings to the gauge
bosons. The hadronic $\tau$ decay modes constitute an
ideal  tool for  studying low--energy effects of the strong
interactions in  very clean conditions; a well--known example is the
precise determination of the QCD coupling from
$\tau$--decay data. 
New physics phenomena, such as a non-zero $m_{\nu_\tau}$ or 
violations of (flavour / CP) conservation laws
can also be searched for with $\tau$ decays.}

\vspace*{0.2cm}
\normalsize\baselineskip=15pt
\setcounter{footnote}{0}
\renewcommand{\thefootnote}{\alph{footnote}}

\section{INTRODUCTION}
\label{sec:introduction}

The $\tau$ lepton is a member of
the third generation which decays into particles belonging to the first
and second ones.
Thus, $\tau$ physics could provide some
clues to the puzzle of the recurring families of leptons and quarks.
In fact, one na\"{\i}vely expects the heavier fermions to be more sensitive to
whatever dynamics is responsible for the fermion--mass generation.

The pure leptonic or semileptonic character of $\tau$  decays
provides a clean laboratory to test the structure of the weak
currents  and the universality of their couplings to the gauge bosons.
Moreover, the  $\tau$ is
the only known lepton massive enough to  decay  into  hadrons;
its  semileptonic decays are then an ideal tool for studying
strong interaction effects in  very clean conditions.

 Since its discovery \cite{PE:75} in 1975 at the SPEAR $e^+ e^-$ 
storage ring, the $\tau$ lepton has been a subject of extensive 
experimental study
\cite{PE:80,HP:88,BS:88,GP:88,KI:88,PI:90,PI:92,PE:92,RI:92,WS:93,GP:96,PI:96}. 
However, it has been during the last few years
when $\tau$ physics has reached its maturity level.
The very clean sample of boosted $\tau^+\tau^-$ events
accumulated at the $Z$ peak, together with the large statistics
collected in the $\Upsilon$ region, have not only considerably improved
the statistical accuracy of the $\tau$ measurements 
but, more importantly, have brought a new
level of systematic understanding.
%
%
%
%
Many of the small ($\sim 2\sigma$) discrepancies which were plaguing
before \cite{PI:92} the $\tau$ data have been already resolved, 
allowing now to make sensible tests of the $\tau$ properties.
The improved quality of the data has motivated a growing interest
on the $\tau$ particle, reflected
in a series of workshops \cite{OR:90,OH:92,MO:94,CO:96}
devoted entirely to the $\tau$.

On the theoretical side, a lot of effort has been invested recently
to improve our understanding of the $\tau$ dynamics.
The basic $\tau$ properties were already known, before its actual
discovery \cite{PE:75}, thanks to the pioneering paper of Tsai \cite{TS:71}.
The detailed study of higher--order electroweak corrections and 
QCD contributions, performed during the last few years, has promoted
the physics of the $\tau$ lepton to the level of precision tests.
There is now an ample
recognition among the physics community of the unique properties
of the $\tau$ for testing the Standard Model, both
in the electroweak and the strong sectors.

All experimental results obtained so far confirm the Standard Model
scenario, in which the $\tau$ is a sequential lepton with its own
quantum number and associated neutrino. With the increased sensitivities
achieved recently, interesting limits on possible new physics contributions
to the $\tau$ decay amplitudes start to emerge.
In the following, the present knowledge on the $\tau$ lepton
is analyzed. Rather than given a detailed
review of experimental results, the emphasis is put on the physics
which can be investigated with the $\tau$ data. Exhaustive information
on more experimental aspects can be found in 
Refs.~\citenum{CO:96} and \citenum{PDG:96}.

\setcounter{equation}{0}
\section{CHARGED--CURRENT UNIVERSALITY}
\label{sec:universality}
 
%
\begin{figure}[htb]
\centerline{\epsfxsize =8cm \epsfbox{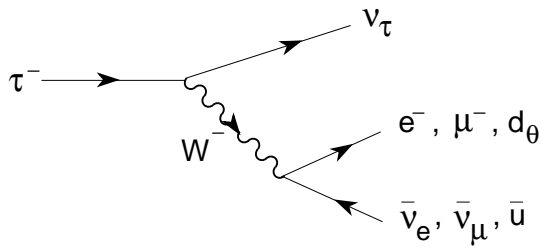}}
\caption{Feynman diagram for the decay of the $\tau$ lepton.}
\label{fig:diagram}
\end{figure}
%

Within the Standard Model, the $\tau$ lepton decays via the 
$W$--emission diagram shown in Figure~\ref{fig:diagram}. 
Since the $W$ coupling to the charged current
is of universal strength,
\bel{eq:L_cc}
\cL_{cc} \, = \, {g \over 2 \sqrt{2}} \, W_{\mu}^\dagger \,
   \left\{ \sum_l \bar{\nu}_l
      \gamma^{\mu} (1 - \gamma_5) l  \, + \, \bar{u} \gamma^{\mu}
      (1 - \gamma_5)
       d_{\theta} \right\} \, + \, \mbox{\rm h.c.}\, ,
\ee
there are five equal contributions
(if final masses and gluonic corrections
are neglected) to the $\tau$ decay width.
Two of them correspond to the leptonic
decay modes $\tau^- \rightarrow \nu_{\tau} e^- \bar{\nu}_e$  and
$\tau^- \rightarrow \nu_{\tau} \mu^- \bar{\nu}_{\mu}$ , while the other
three are associated with the three possible
colours of the quark--antiquark
pair in the $\tau^- \rightarrow \nu_{\tau} d_{\theta} \bar{u}$
decay mode
($d_{\theta} \equiv \cos{\theta_C} d + \sin{\theta_C} s$).
Hence, the branching ratios for the different channels
are expected to be approximately:
\begin{eqnarray}
\label{eq:naive}
 B_l &  \equiv &
    \mbox{\rm Br}(\tau^- \rightarrow \nu_{\tau}l^- \bar{\nu}_l) \,
    \simeq \, {1 \over 5} \, = \,
    20\% \qquad  (l=e, \mu) ,\\
 R_{\tau } &
\equiv & { \Gamma (\tau^- \rightarrow \nu_{\tau} + 
  \mbox{\rm hadrons}) \over
        \Gamma (\tau^- \rightarrow \nu_{\tau} e^- \bar{\nu}_e ) }
     \, \simeq \, N_C \, = \, 3 \, ,
\end{eqnarray}
which should be compared with the present experimental averages
\cite{CO:96,PDG:96} in Table 1.
The agreement is fairly good. Notice that the measured $\tau$
hadronic width
provides strong evidence for the colour degree of freedom.
We will discuss later whether the QCD dynamics
is able to explain the (20\%) difference between the
measured value of $R_{\tau}$ and the lowest--order prediction
$R_{\tau} = N_C$.
 
%
\begin{table}[thb]
\caption{Average values \protect\cite{CO:96,PDG:96}
of some basic $\tau$ parameters. 
$h^-$ stands for either $\pi^-$ or $K^-$.}
\label{tab:parameters}
\centering\vspace{0.2cm}
\begin{tabular}{|c|c|}
\hline 
$m_\tau$ & $(1777.00^{+0.30}_{-0.27})$ MeV \\
$\tau_\tau$ & $(290.21\pm 1.15)$ fs \\
Br($\tau^-\to\nu_{\tau}e^-\bar{\nu}_e$)   
   & $(17.786\pm 0.072)\% $ \\
Br($\tau^-\to\nu_{\tau}\mu^-\bar{\nu}_\mu$)   
   & $(17.317\pm 0.078)\% $ \\
$R_\tau$ & $3.649\pm 0.014$ \\
Br($\tau^-\to\nu_\tau\pi^-$) & $(11.01\pm 0.11)\% $ \\
Br($\tau^-\to\nu_\tau K^-$) & $(0.692\pm 0.028)\% $ \\
Br($\tau^-\to\nu_\tau h^-$) & $(11.70\pm 0.11)\% $ \\
\hline
\end{tabular}
\end{table}
%

The leptonic decays 
$\tau^-\to l^-\bar\nu_l\nu_\tau$ ($l=e,\mu$)
are theoretically understood at the level of the electroweak
radiative corrections \cite{MS:88}.
Within the Standard Model  (neutrinos are assumed to be massless),
\begin{equation}
\label{eq:leptonic}
\Gamma_{\tau\to l} \, \equiv \,
\Gamma (\tau^- \rightarrow \nu_{\tau} l^- \bar{\nu}_l)  \, = \,
  {G_F^2 m_{\tau}^5 \over 192 \pi^3} \, 
  f\!\left({m_l^2 \over m_{\tau}^2}\right) \, r_{EW},
\end{equation}
where $f(x) = 1 - 8 x + 8 x^3 - x^4 - 12 x^2 \log{x}$.
The factor $r_{EW}$ takes into account radiative corrections not
included in the
Fermi coupling constant $G_F$, and the non-local structure of the
$W$ propagator; these effects \cite{MS:88} are quite small
[$\alpha(m_\tau) = 1 / 133.3 $]:
\bel{eq:e_EW}
r_{EW} \, = \, \left[1 + {\alpha(m_\tau) \over 2 \pi } 
\left({25 \over 4}  - \pi^2 \right) \right] \, 
\left[ 1 + { 3 \over 5 } {m_\tau^2 \over M_W^2}
  - 2 {m_l^2 \over M_W^2} \right]
\, = \,  0.9960 \, . 
\ee

Using the value of $G_F$ measured in $\mu$ decay,
Eq.~\eqn{eq:leptonic} provides a relation between the
$\tau$ lifetime and the leptonic branching ratios:
\bel{relation}
B_e \, = \, {B_\mu \over 0.972564\pm 0.000010} \, =
{ \tau_{\tau} \over (1632.1 \pm 1.4) \times 10^{-15}\, {\rm s} } \, .
\ee
The quoted errors reflect the present uncertainty of $0.3$ MeV
in the value of $m_\tau$.

\begin{figure}[bht]
\vspace{0.5cm}
\centerline{\epsfxsize =9.5cm \epsfbox{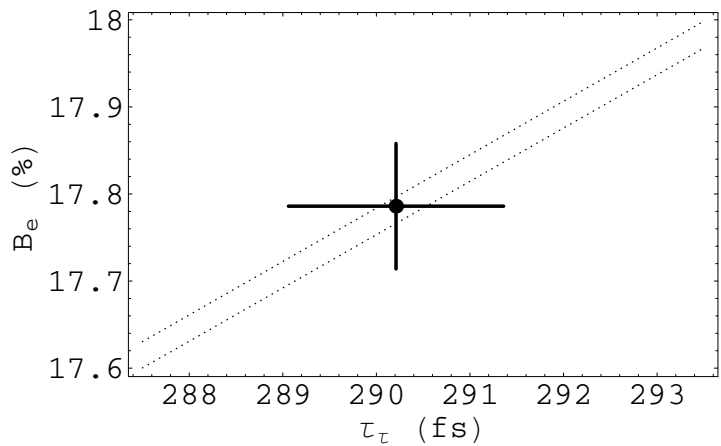}}
\caption{Relation between $B_e$ and $\tau_\tau$. The dotted
band corresponds to Eq.~(\protect\ref{relation}).}
\label{fig:BeLife}
\end{figure}

The predicted value of $B_\mu/B_e$ is in perfect agreement with the 
measured ratio
$B_\mu/B_e = 0.974 \pm 0.006$.  As shown in
Figure~\ref{fig:BeLife}, the relation between $B_e$ and
$\tau_\tau$ is also well satisfied by the present data. Notice, that this
relation is very sensitive to the value of the $\tau$ mass
[$\Gamma_{\tau\to l}\propto m_\tau^5$]. The most recent measurements of
$\tau_\tau$, $B_e$ and $m_\tau$ have consistently moved the world averages
in the correct direction, eliminating the previous ($\sim 2\sigma$)
disagreement \cite{PI:92}. 
The experimental precision (0.4\%) is already approaching the
level where a possible non-zero $\nu_\tau$ mass could become relevant;
the present bound \cite{ALEPH:95a}
$m_{\nu_\tau}< 24$ MeV (95\% CL) only guarantees that such 
effect\footnote{
The preliminary ALEPH bound \protect\cite{Passalacqua},
$m_{\nu_\tau}< 18.2$ MeV (95\% CL), implies a correction smaller than
0.08\% .} \
is below 0.14\%.

These measurements can be used to test the universality of
the $W$ couplings to the leptonic charged currents.
The $B_\mu/B_e$ ratio constraints $|g_\mu/g_e|$, while the
$B_e/\tau_\tau$ relation provides information on $|g_\tau/g_\mu|$.
The present results are shown in Tables \ref{tab:univme} and
\ref{tab:univtm}, together with the values obtained from the
$\pi$--decay ratio \cite{BR:92}
$R_{\pi\to e/\mu}\equiv\Gamma(\pi^-\to e^-\bar\nu_e)/
\Gamma(\pi^-\to\mu^-\bar\nu_\mu)$,
and from the comparison of the $\sigma\cdot B$ partial production
cross--sections for the various $W^-\to l^-\bar\nu_l$ decay
modes at the $p$-$\bar p$ colliders \cite{UA1:89,UA2:91,CDF:92}. 

\begin{table}[bt]
\caption{Present constraints \protect\cite{PI:96} on $|g_\mu/g_e|$.}
\label{tab:univme}
\centering\vspace{0.2cm}
\begin{tabular}{|c|c|c|c|}
\hline
& $B_\mu/B_e$  & $R_{\pi\to e/\mu}$ & $\sigma\cdot B_{W\to\mu/e}$ 
\\ \hline
$|g_\mu/g_e|$ & $1.0005\pm 0.0030$ & $1.0017\pm 0.0015$ &
$1.01\pm 0.04$
\\ \hline
\end{tabular}
\end{table}
\begin{table}[bt]
\caption{Present constraints \protect\cite{PI:96} on $|g_\tau/g_\mu|$.}
\label{tab:univtm}
\centering\vspace{0.2cm}
\begin{tabular}{|c|c|c|c|c|}
\hline
& $B_e\tau_\mu/\tau_\tau$ & $R_{\tau/\pi}$ & $R_{\tau/K}$ &
$\sigma\cdot B_{W\to\tau/\mu}$ 
\\ \hline
$|g_\tau/g_\mu|$ & $1.0001\pm 0.0029$ & $1.005\pm 0.005$ &
$0.984\pm 0.020$ & $0.99\pm 0.05$ 
\\ \hline
\end{tabular}
\end{table}

The decay modes $\tau^-\to\nu_\tau\pi^-$ and $\tau^-\to\nu_\tau K^-$ 
can also be used to test universality through the ratios
\beqn\label{eq:R_tp}
R_{\tau/\pi} & \!\!\!\equiv &\!\!\!
 {\Gamma(\tau^-\to\nu_\tau\pi^-) \over
 \Gamma(\pi^-\to \mu^-\bar\nu_\mu)} =
\Big\vert {g_\tau\over g_\mu}\Big\vert^2
{m_\tau^3\over 2 m_\pi m_\mu^2}
{(1-m_\pi^2/ m_\tau^2)^2\over
 (1-m_\mu^2/ m_\pi^2)^2} 
\left( 1 + \delta R_{\tau/\pi}\right) , \qquad
\\ \label{eq:R_tk} 
R_{\tau/K} &\!\!\! \equiv &\!\!\! {\Gamma(\tau^-\to\nu_\tau K^-) \over
 \Gamma(K^-\to \mu^-\bar\nu_\mu)} =
\Big\vert {g_\tau\over g_\mu}\Big\vert^2
{m_\tau^3\over 2 m_K m_\mu^2}
{(1-m_K^2/m_\tau^2)^2\over
(1-m_\mu^2/ m_K^2)^2} 
\left( 1 + \delta R_{\tau/K}\right) , \qquad
\eeqn
where the dependence on the hadronic matrix elements (the so-called
decay constants $f_{\pi,K}$) factors out.
Owing to the different energy scales involved, the radiative
corrections to the $\tau^-\to\nu_\tau\pi^-/K^-$ amplitudes
are however not the same than the corresponding effects in
$\pi^-/K^-\to\mu^-\bar\nu_\mu$. The size of the relative
correction has been estimated \cite{MS:93,DF:94} to be:
\bel{eq:dR_tp_tk}
\delta R_{\tau/\pi} = (0.16\pm 0.14)\% \ ,    \qquad\qquad
\delta R_{\tau/K} = (0.90\pm 0.22)\%  \ .
\ee
Using these numbers, the measured $\tau^-\to\pi^-\nu_\tau$
and $\tau^-\to K^-\nu_\tau$ decay rates imply the
$|g_\tau/g_\mu|$ ratios given in Table~\ref{tab:univtm}.
The inclusive sum of both decay modes
provides a
slightly more accurate determination:
$|g_\tau/g_\mu| = 1.004\pm 0.005$.

The present data verify the universality of the leptonic
charged--current couplings to the 0.15\% ($e/\mu$) and 0.30\%
($\tau/\mu$) level. The precision of the most recent
$\tau$--decay measurements is becoming competitive with the 
more accurate $\pi$--decay determination. 
It is important to realize the complementarity of the
different universality tests. 
The pure leptonic decay modes probe
the charged--current couplings of a transverse $W$. In contrast,
the decays $\pi/K\to l\bar\nu$ and $\tau\to\nu_\tau\pi/K$ are only
sensitive to the spin--0 piece of the charged current; thus,
they could unveil the presence of possible scalar--exchange
contributions with Yukawa--like couplings proportional to some
power of the charged--lepton mass.
One can easily imagine new--physics scenarios which would modify 
differently the two types of leptonic couplings \cite{MA:94}. 
For instance,
in the usual two--Higgs doublet model, charged--scalar exchange
generates a correction to the ratio $B_\mu/B_e$, but 
$R_{\pi\to e/\mu}$ remains unaffected.
Similarly, lepton mixing between the $\nu_\tau$ and an hypothetical
heavy neutrino would not modify the ratios  $B_\mu/B_e$ and
$R_{\pi\to e/\mu}$, but would certainly correct the relation between
$B_l$ and the $\tau$ lifetime.
 
\section{LORENTZ STRUCTURE OF THE CHARGED CURRENT}
\label{sec:current}
\setcounter{equation}{0} 

Let us consider the leptonic decays $l^-\to\nu_l l'^-\bar\nu_{l'}$, 
where the lepton pair ($l$, $l^\prime $)
may be ($\mu$, $e$), ($\tau$, $e$), or ($\tau$, $\mu$). 
The most general, local, derivative--free, lepton--number conserving, 
four--lepton interaction Hamiltonian, 
consistent with locality and Lorentz invariance
\cite{MI:50,BM:57,KS:57,SCH:83,FGJ:86,FG:93,PS:95},
\be
{\cal H} \, =\,  4\, \frac{G_{l'l}}{\sqrt{2}}\,
\sum_{n,\epsilon,\omega}
g^n_{\epsilon\omega}
\left[ \overline{l'_\epsilon} 
\Gamma^n {(\nu_{l'})}_\sigma \right]\, 
\left[ \overline{({\nu_l})_\lambda} \Gamma_n 
	l_\omega \right]\ ,
\label{eq:hamiltonian}
\ee
contains ten complex coupling constants or, since a common phase is
arbitrary, nineteen independent real parameters
which could be different for each leptonic decay.
The subindices $\epsilon , \omega , \sigma, \lambda$ label 
the chiralities (left--handed,
right--handed)  of the  corresponding  fermions, and $n$ the
type of interaction: 
scalar ($I$), vector ($\gamma^\mu$), tensor 
($\sigma^{\mu\nu}/\sqrt{2}$).
For given $n, \epsilon ,
\omega $, the neutrino chiralities $\sigma $ and $\lambda$
are uniquely determined.

Taking out a common factor $G_{l'l}$, which is determined by the total
decay rate, the coupling constants $g^n_{\epsilon\omega}$
are normalized to \cite{FGJ:86}
\beqn\label{eq:normalization}
1 &\!\!\! = &\!\!\!
{1\over 4} \,\left( |g^S_{RR}|^2 + |g^S_{RL}|^2
    + |g^S_{LR}|^2 + |g^S_{LL}|^2 \right)
  \, + \, 3 \,\left( |g^T_{RL}|^2 + |g^T_{LR}|^2 \right) 
\no \\ & &\!\!\!\! \mbox{}
+ \left(
   |g^V_{RR}|^2 + |g^V_{RL}|^2 + |g^V_{LR}|^2 + |g^V_{LL}|^2 \right)
\, .
\eeqn
In the Standard Model, $g^V_{LL}  = 1$  and all other
$g^n_{\epsilon\omega} = 0 $.

For an initial lepton polarization ${\cal P}_l$,
the final charged lepton distribution in the decaying lepton 
rest frame
is usually parametrized in the form  \cite{BM:57,KS:57}
\be\label{eq:spectrum}
{d^2\Gamma_{l\to l'} \over dx\, d\cos\theta} =
{m_l\,\omega^4 \over 2\pi^3} G_{l'l}^2 \sqrt{x^2-x_0^2}
\left\{ F(x) - 
{\xi\over 3}\, {\cal P}_l\,\sqrt{x^2-x_0^2}
\,\cos{\theta}\, A(x)\right\} ,
\ee
where $\theta$ is the angle between the $l^-$ spin and the
final charged--lepton momentum,
$\, \omega \equiv (m_l^2 + m_{l'}^2)/2 m_l \, $
is the maximum $l'^-$ energy for massless neutrinos, $x \equiv E_{l'^-} /
\omega$ is the reduced energy, $x_0\equiv m_{l'}/\omega$
and
\beqn\label{eq:Fx_Ax_def}
F(x)  &\!\!\! = &\!\!\! 
  x (1 - x) + {2\over 9} \rho 
 \left(4 x^2 - 3 x - x_0^2 \right) 
+  \eta\, x_0 (1-x)
\, , \no\\
A(x) &\!\!\! = &\!\!\! 
 1 - x   + {2\over 3}  \delta \left( 4 x - 4 + \sqrt{1-x_0^2}  
\right)  \, .
\eeqn

For unpolarized $l's$, the distribution is characterized by
the so-called Michel \cite{MI:50} parameter $\rho$
and the low--energy parameter $\eta$. Two more parameters, $\xi$
and $\delta$, can be determined when the initial lepton polarization is known.
If the polarization of the final charged lepton is also measured,
5 additional independent parameters \cite{PDG:96}  
($\xi'$, $\xi''$, $\eta''$, $\alpha'$, $\beta'$)
appear. 

For massless neutrinos, the total decay rate is given by \cite{PS:95}
\be\label{eq:gamma}
\Gamma_{l\to l'} \, = \, {\widehat G_{l'l}^2 m_l^5\over 192 \pi^3}\,
f\!\left({m_{l'}^2\over m_l^2}\right) 
 r_{\mbox{\rms EW}}
\, ,
\ee
where
\be\label{eq:Ghat_def}
\widehat G_{l'l} \,\equiv\, G_{l'l} \,
\sqrt{1 + 4\,\eta\, {m_{l'}\over m_l}\,
{g\!\left( m_{l'}^2/ m_l^2 \right)\over  
f\!\left( m_{l'}^2/ m_l^2 \right)}}
\, ,
\ee
$g(z) = 1 + 9 z - 9 z^2 - z^3 + 6 z (1+z) \ln{z}$,
and the Standard Model
radiative correction\footnote{
Since we assume that the Standard Model provides the dominant 
contribution
to the decay rate, any additional higher--order correction
beyond the effective Hamiltonian 
(\protect\ref{eq:hamiltonian})
would be a sub\-leading effect.} \
$r_{\mbox{\rms EW}}$ has been included.
Thus, the normalization $G_{e\mu}$ corresponds to the Fermi coupling 
$G_F$, measured in $\mu$ decay.
The $B_\mu/B_e$ and $B_e\tau_\mu/\tau_\tau$
universality tests, discussed in the previous section,
actually prove the ratios
$|\widehat G_{\mu\tau}/\widehat G_{e\tau}|$
and $|\widehat G_{e\tau}/\widehat G_{e\mu}|$, respectively.
An important point, emphatically stressed by
Fetscher and Gerber \cite{FG:93}, concerns the extraction
of $G_{e \mu}$, whose uncertainty is dominated
by the uncertainty in $\eta_{\mu\to e}$. 

In terms of the $g_{\epsilon\omega}^n$
couplings, the shape parameters in Eqs.~\eqn{eq:spectrum} and
\eqn{eq:Fx_Ax_def} are:
\beqn\label{eq:michel}
\rho  & \!\!\! = & \!\!\! 
{3\over 4} (\beta^+ + \beta^-) + (\gamma^+ + \gamma^-) \, ,
\no\\
\xi & \!\!\! = & \!\!\! 3 (\alpha^- - \alpha^+) + (\beta^- - \beta^+)
  + {7\over 3} (\gamma^+ - \gamma^-) \, ,
\no\\
\xi\delta & \!\!\! = & \!\!\!
 {3\over 4} (\beta^- - \beta^+) + (\gamma^+ - \gamma^-) \, ,
\\
\eta 
& \!\!\! = & \!\!\!
\frac{1}{2} \mbox{\rm Re}\left[
g^V_{LL} g^{S\ast}_{RR} + g^V_{RR}  g^{S\ast}_{LL}
+ g^V_{LR} \left(g^{S\ast}_{RL} + 6 g^{T\ast}_{RL}\right) 
+ g^V_{RL} \left(g^{S\ast}_{LR} + 6 g^{T\ast}_{LR}\right) 
\right] ,
\no
\eeqn
where \cite{Rouge}
\beqn\label{eq:abg_def}
\lefteqn{\alpha^+ \equiv  
{|g^V_{RL}|}^2 + {1\over 16} {|g^S_{RL} + 6 g^T_{RL}|}^2
\, , } && \qquad\qquad\qquad\qquad\qquad\qquad\qquad
\alpha^- \equiv  
{|g^V_{LR}|}^2 + {1\over 16} {|g^S_{LR} + 6 g^T_{LR}|}^2
\, , 
\no\\
\lefteqn{\beta^+ \equiv 
 {|g^V_{RR}|}^2 + {1\over 4} {|g^S_{RR}|}^2
\, , } && \qquad\qquad\qquad\qquad\qquad\qquad\qquad
\beta^- \equiv 
 {|g^V_{LL}|}^2 + {1\over 4} {|g^S_{LL}|}^2
\, ,
\\
\lefteqn{\gamma^+ \equiv 
 {3\over 16} {|g^S_{RL} - 2 g^T_{RL}|}^2
\, , } && \qquad\qquad\qquad\qquad\qquad\qquad\qquad
\gamma^- \equiv 
 {3\over 16} {|g^S_{LR} - 2 g^T_{LR}|}^2
\, ,
\no
\eeqn
are positive--definite combinations of decay constants, corresponding to 
a final right--handed ($\alpha^+,\beta^+,\gamma^+$) or
left--handed ($\alpha^-,\beta^-,\gamma^-$)
lepton.
In the Standard Model, 
$\rho = \delta = 3/4$, 
$\eta = \eta'' = \alpha' = \beta' = 0 $ and 
$\xi = \xi' = \xi'' = 1 $.

The normalization constraint \eqn{eq:normalization} is equivalent to
$\alpha^+ + \alpha^- + \beta^+ + \beta^- + \gamma^+ + \gamma^- = 1$.
It is convenient to introduce \cite{FGJ:86} the probabilities
$Q_{\epsilon\omega}$ for the
decay of an $\omega$--handed $l^-$
into an $\epsilon$--handed 
daughter lepton,
\begin{eqnarray}\label{eq:Q_LL}
Q_{LL}
&\!\!\! = &\!\!\!  \beta^- =
{1 \over 4} |g^S_{LL}|^2 \! +  |g^V_{LL}|^2 
 = {1 \over 4}\left(
-3 +{16\over 3}\rho -{1\over 3}\xi +{16\over 9}\xi\delta +\xi'+\xi''
\right)\! , \quad\;\;\no\\ 
Q_{RR} 
&\!\!\! = &\!\!\!  \beta^+ =
{1 \over 4} |g^S_{RR}|^2 \! + \! |g^V_{RR}|^2 
 =  {1 \over 4}\left(
-3 +{16\over 3}\rho +{1\over 3}\xi -{16\over 9}\xi\delta -\xi'+\xi''
\right)\!  , \\ 
Q_{LR} 
&\!\!\! = &\!\!\! \alpha^- + \gamma^- =
{1 \over 4} |g^S_{LR}|^2 \! + \!  |g^V_{LR}|^2
            \! + \!   3 |g^T_{LR}|^2  
 = {1 \over 4}\left(
5 -{16\over 3}\rho +{1\over 3}\xi -{16\over 9}\xi\delta +\xi'-\xi''
\right)\! , \no\\ 
Q_{RL}
&\!\!\! = &\!\!\! \alpha^+ + \gamma^+ =
{1 \over 4} |g^S_{RL}|^2  \! + \!  |g^V_{RL}|^2
            \! + \!  3 |g^T_{RL}|^2  
= {1 \over 4}\left(
5 -{16\over 3}\rho -{1\over 3}\xi +{16\over 9}\xi\delta -\xi'-\xi''
\right)\! . \no
\end{eqnarray}
Upper bounds on any of these (positive--semidefinite) probabilities 
translate into corresponding limits for all couplings with the 
given chiralities.

\begin{figure}[bth]
\vfill
\centerline{
\begin{minipage}[t]{.47\linewidth}\centering
\centerline{\epsfxsize =7.5cm \epsfbox{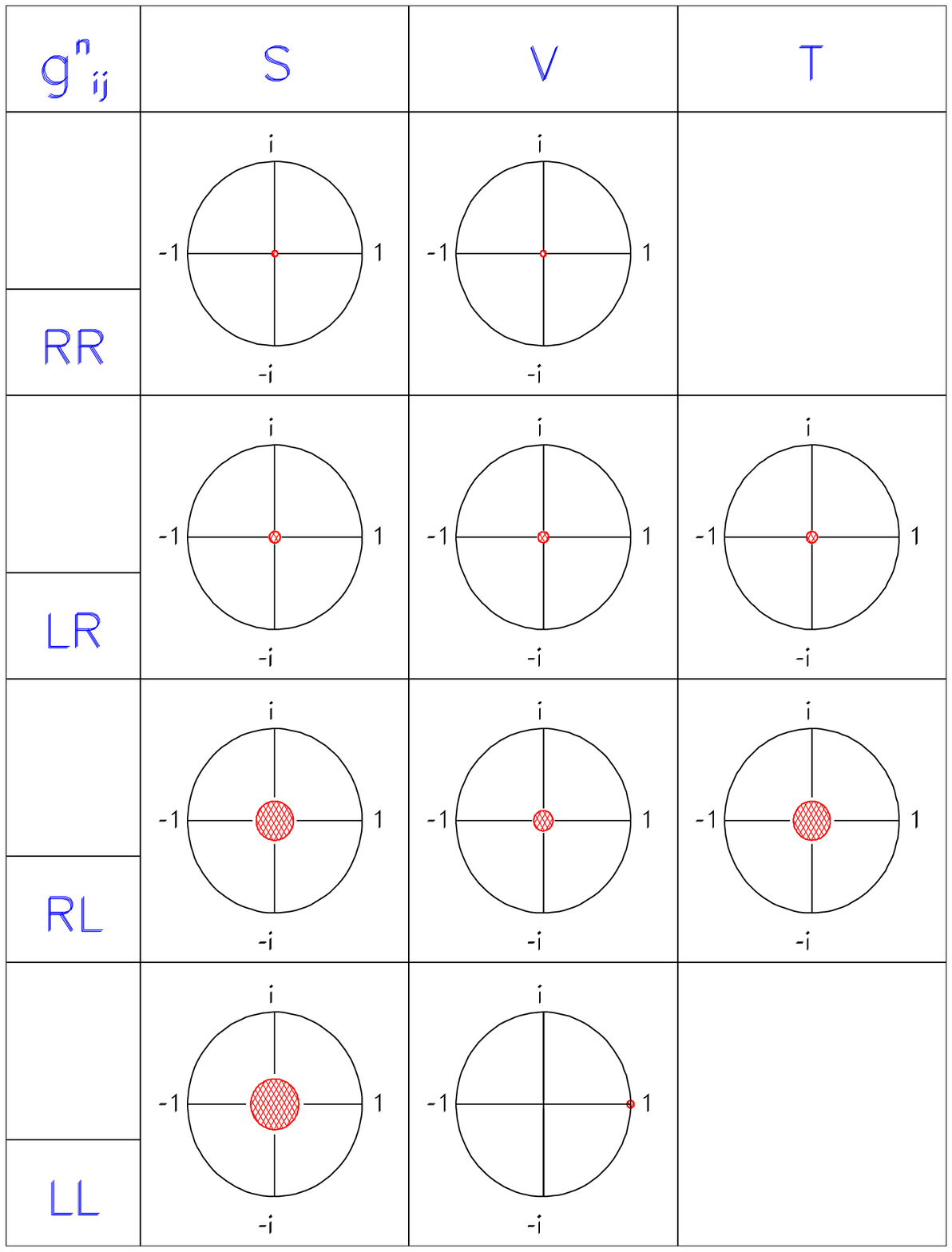}}
\caption{90\% CL experimental limits  \protect\cite{PDG:96}
for the normalized $\mu$--decay couplings
$g'^n_{\epsilon\omega }\equiv g^n_{\epsilon\omega }/ N^n$,
where
$N^n \equiv \protect\mbox{\rm max}(|g^n_{\epsilon\omega }|) =2$,
1, $1/\protect\sqrt{3} $ for $n =$ S, V, T.
(Taken from Ref.~\protect\citenum{LR:95}).}
\label{fig:mu_couplings}
\end{minipage}
\hspace{0.67cm}
\begin{minipage}[t]{.47\linewidth}\centering
\centerline{\epsfxsize =7.5cm \epsfbox{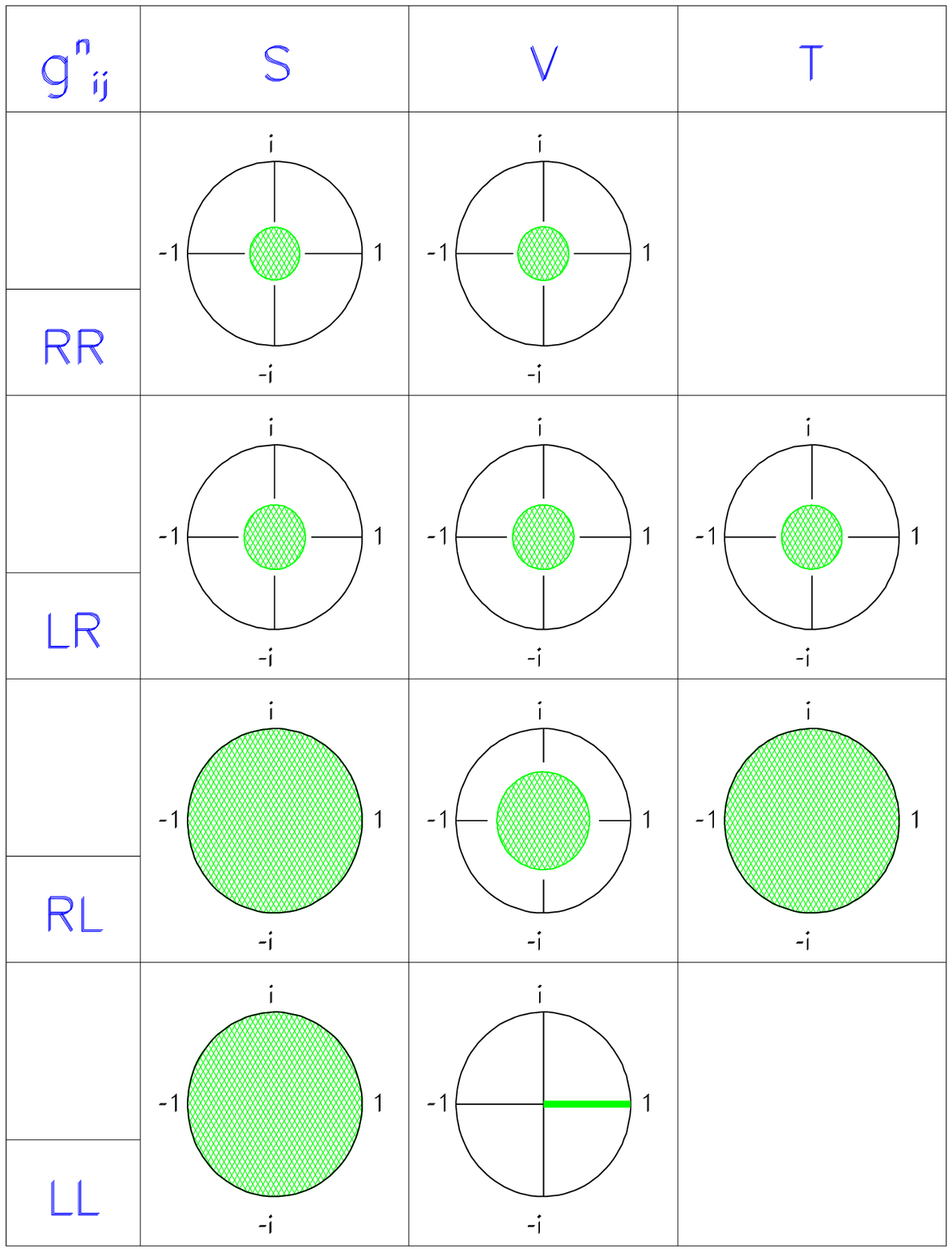}}
\caption{90\% CL experimental limits \cite{PI:96}
for the normalized $\tau$--decay couplings
$g'^n_{\epsilon\omega }\equiv g^n_{\epsilon\omega }/ N^n$,
assuming $e/\mu$ universality.
\label{fig:tau_couplings}}
\end{minipage}
}
\vfill
\end{figure}

For $\mu$ decay, where precise measurements of the polarizations of
both $\mu$ and $e$ have been performed, there exist \cite{FGJ:86}
upper bounds on $Q_{RR}$, $Q_{LR}$ and $Q_{RL}$, and a lower bound
on $Q_{LL}$. They imply corresponding upper bounds on the 8
couplings $|g^n_{RR}|$, $|g^n_{LR}|$ and $|g^n_{RL}|$.
The measurements of the $\mu^-$ and the $e^-$ do not allow to
determine $|g^S_{LL}|$ and $|g^V_{LL}|$ separately \cite{FGJ:86,JA:66}.
Nevertheless, since the helicity of the $\nu_\mu$ in pion decay is
experimentally known \cite{RO:82}
to be $-1$, a lower limit on $|g^V_{LL}|$ is
obtained \cite{FGJ:86} from the inverse muon decay
$\nu_\mu e^-\to\mu^-\nu_e$.
The present (90\% CL) bounds \cite{PDG:96} 
on the $\mu$--decay couplings are shown in Figure~\ref{fig:mu_couplings}.
These limits show nicely 
that the bulk of the $\mu$--decay transition amplitude is indeed of
the predicted V$-$A type.

The experimental analysis of the $\tau$--decay parameters is 
necessarily
different from the one applied to the muon, because of the much
shorter $\tau$ lifetime.
The measurement of the $\tau$ polarization and the parameters
$\xi$ and $\delta$ is still possible due to the fact that the spins
of the $\tau^+\tau^-$ pair produced in $e^+e^-$ annihilation 
are strongly correlated
\cite{TS:71,KST:73,PS:77,KW:84,GO:89,NE:91,GN:91,FE:90,BPR:91,ABGPR:92,DDDR:93}.
Another possibility is to use
the beam polarization, as done by SLD \cite{SLD:97}.
However,
the polarization of the charged lepton emitted in the $\tau$ decay
has never been measured. In principle, this could be done
for the decay $\tau^-\to\mu^-\bar\nu_\mu\nu_\tau$ by stopping the
muons and detecting their decay products \cite{FE:90}.
An alternative method would be \cite{SV:96} to use 
the radiative decays
$\tau\to l^-\bar\nu_l\nu_\tau\gamma$ ($l=e,\mu$), 
since the
distribution of the photons emitted by the daughter lepton is
sensitive to the lepton polarization. 
The measurement of the inverse decay $\nu_\tau l^-\to\tau^-\nu_l$
looks far out of reach.

The present experimental status \cite{CO:96} on the $\tau$--decay 
Michel parameters is shown in Table~\ref{tab:tau_michel}.
For comparison, the values measured in $\mu$ decay \cite{PDG:96}
are also given.
The improved accuracy of the most recent experimental analyses
has brought an enhanced sensitivity to the different shape parameters,
allowing the first measurements 
\cite{CO:96,SLD:97,evans,chadha,ALEPH:95b}
of $\eta_{\tau\to\mu}$ 
$\xi_{\tau\to e}$, $\xi_{\tau\to\mu}$, $(\xi\delta)_{\tau\to e}$ and 
$(\xi\delta)_{\tau\to\mu}$, 
without any $e/\mu$ universality assumption.

\begin{table}[bt]   
\caption{World average \protect\cite{CO:96,PDG:96}
Michel parameters. 
The last column ($\tau\to l$) assumes identical couplings
for $l=e,\mu$.
$\xi_{\mu\to e}$ refers to the product $\xi_{\mu\to e}\cP_\mu$,
where $\cP_\mu\approx 1$ is the longitudinal polarization
of the $\mu$ from $\pi$ decay.}
\label{tab:tau_michel}
\centering\vspace{0.2cm}
\begin{tabular}{|c|c|c|c|c|}
\hline
& $\mu\to e$ & $\tau\to\mu$ & $\tau\to e$ & $\tau\to l$ 
\\ \hline
$\rho$ & $0.7518\pm 0.0026$ & $0.733\pm 0.031$ & $0.734\pm 0.016$ & 
$0.741\pm 0.014$ 
\\
$\eta$ & $-0.007\pm 0.013\phantom{-}$ & $-0.04\pm 0.20\phantom{-}$ & --- & 
$0.047\pm 0.076$
\\
$\xi$ & $1.0027\pm 0.0085$ & $1.19\pm 0.18$ & $1.09\pm 0.16$ & 
$1.04\pm 0.09$ 
\\
$\xi\delta$ & $0.7506\pm 0.0074$ & $0.73\pm 0.11$ & $ 0.80\pm 0.18$ & 
$ 0.73\pm 0.07$ 
\\ \hline
\end{tabular}
\end{table}

The determination of the $\tau$--polarization parameters
allows us to bound the total probability for the decay of
a right--handed $\tau$ \cite{FE:90},
\be\label{eq:Q_R}
Q_{\tau_R} \equiv Q_{RR} + Q_{LR} 
= \frac{1}{2}\, \left[ 1 + \frac{\xi}{3} - \frac{16}{9} 
(\xi\delta)\right]
\; .
\ee
One finds \cite{PI:96}
(ignoring possible correlations among the measurements):
\begin{eqnarray}
Q_{\tau_R}^{\tau\to\mu} &\!\!\! =&\!\!\! \phantom{-}0.05\pm 0.10 \; 
< \, 0.20 \quad (90\%\;\mbox{\rm CL})\, , \no\\
Q_{\tau_R}^{\tau\to e} &\!\!\! =&\!\!\! -0.03\pm 0.16 \; 
< \, 0.25 \quad (90\%\;\mbox{\rm CL})\, , \\
Q_{\tau_R}^{\tau\to l} &\!\!\! =&\!\!\! \phantom{-}0.02\pm 0.06 \;
< \, 0.12 \quad (90\%\;\mbox{\rm CL})\, , \no
\end{eqnarray}
where the last value refers to the $\tau$ decay into either $l=e$ or $\mu$,
assuming identical $e$/$\mu$ couplings.
Since these probabilities are positive--semidefinite quantities, they imply
corresponding limits on all
$|g^n_{RR}|$ and $|g^n_{LR}|$ couplings. 

A measurement of the final lepton polarization could be even more efficient,
since the total probability for the decay into a right--handed lepton
depends on a single Michel parameter:
\bel{eq:Qxi}
Q_{l'_R} \equiv Q_{RR} +  Q_{RL}
= {1\over 2} ( 1 -\xi') \, .
\ee
Thus, a single polarization measurement could bound the five RR and RL
complex couplings.

Another useful positive--semidefinite quantity is \cite{LR:95}
\bel{eq:rxd}
\rho - \xi\delta = {3\over 2} \beta^+ + 2 \gamma^- \, ,
\ee
which provides direct bounds on $|g^V_{RR}|$ and $|g^S_{RR}|$.
A rather weak upper limit on $\gamma^+$ is obtained from the
parameter $\rho$. More stringent is the bound on $\alpha^+$ obtained from
$(1-\rho)$, which is also positive--semidefinite; it implies a corresponding
limit on $|g^V_{RL}|$.

Table~\ref{table:g_tau_bounds} gives \cite{PI:96}
the resulting (90\% CL) bounds on the 
$\tau$--decay couplings.
The relevance of these limits can be better appreciated in 
Figure~\ref{fig:tau_couplings}, 
where $e$/$\mu$ universality has been assumed.

\begin{table}[bth]
\centering
\caption{90\% CL limits
for the $g^n_{\epsilon\omega}$ couplings \protect\cite{PI:96}.}
\label{table:g_tau_bounds}
\vspace{0.2cm}
\begin{tabular}{|l|l|l|l|l|}
\hline
& \hfil $\mu\to e$\hfil &
\hfil $\tau\to\mu$\hfil &\hfil $\tau\to e$ \hfil & 
\hfil $\tau\to l$ \hfil
\\ \hline
$|g^S_{RR}|$  & $< 0.066$ & $< 0.71$ & $< 0.83$ & $< 0.57$
\\
$|g^S_{LR}|$  & $< 0.125$ & $< 0.90$ & $< 1.00$ & $< 0.70$
\\
$|g^S_{RL}|$  & $< 0.424$ & $\leq 2$ & $\leq 2$ & $\leq 2$
\\
$|g^S_{LL}|$  & $< 0.55$  & $\leq 2$ & $\leq 2$ & $\leq 2$
\\ \hline
$|g^V_{RR}|$  & $< 0.033$ & $< 0.36$ & $< 0.42$ & $< 0.29$
\\
$|g^V_{LR}|$  & $< 0.060$ & $< 0.45$ & $< 0.50$ & $< 0.35$
\\
$|g^V_{RL}|$  & $< 0.110$ & $< 0.56$ & $< 0.54$ & $< 0.53$
\\
$|g^V_{LL}|$  & $> 0.96$  & $\leq 1$ & $\leq 1$ & $\leq 1$
\\ \hline
$|g^T_{LR}|$  & $< 0.036$ & $< 0.26$ & $< 0.29$ & $< 0.20$
\\
$|g^T_{RL}|$  & $< 0.122$ & $\leq 1/\sqrt{3}$ & $\leq 1/\sqrt{3}$
              & $\leq 1/\sqrt{3}$
\\ \hline
\end{tabular}
\end{table}

If lepton universality is assumed, 
the leptonic decay ratios $B_\mu/B_e$  and $B_e\tau_\mu/\tau_\tau$
provide limits on the low--energy parameter $\eta$.  
The best sensitivity \cite{ST:94} comes from
$\widehat{G}_{\mu\tau}$,
where the term proportional to $\eta$ is not suppressed by
the small $m_e/m_l$ factor. The measured $B_\mu/B_e$ ratio implies
then:
\be\label{eq:eta_univ}
\eta_{\tau\to l} \, = \, 0.005\pm 0.027  \ .
\ee
This determination is more accurate that the one in 
Table~\ref{tab:tau_michel},
obtained from the shape of the energy distribution,
and is comparable to the value measured in $\mu$ decay.

A non-zero value of $\eta$ would show that there are at least two
different couplings with opposite chiralities for the charged leptons.
Assuming the V$-$A coupling $g_{LL}^V$ to be dominant, the
second one would be \cite{FE:90} a Higgs--type coupling $g^S_{RR}$.
To first order in new physics contributions,
$\eta\approx\mbox{\rm Re}(g^S_{RR})/2$;
Eq.~(\ref{eq:eta_univ}) puts then the (90\% CL) bound:
$-0.08 \, <\mbox{\rm Re}(g^S_{RR}) < 0.10$.

\subsection{Model--Dependent Constraints}

The sensitivity of the present data is not
good enough to get strong constraints from a completely
general analysis of the four--fermion Hamiltonian.
Nevertheless, better limits can be obtained within particular
models, as shown in Tables~\ref{tab:coup_CH} and \ref{tab:W_couplings}.

\begin{table}[thb]
\caption{90\% CL limits 
for the couplings $g^n_{\epsilon\omega}$, assuming that there are no
tensor couplings.}
\label{tab:coup_CH}
\centering\vspace{0.2cm}
\begin{tabular}{|l|l|l|l|l|}
\hline
& \hfil $\mu\to e$\hfil &
\hfil $\tau\to\mu$\hfil &\hfil $\tau\to e$ \hfil & 
\hfil $\tau\to l$ \hfil
\\ \hline
$|g^S_{RR}|$  & $< 0.066$ & $< 0.71$ & $< 0.70$ & $< 0.55$
\\
$|g^S_{LR}|$  & $< 0.125$ & $< 0.71$ & $< 0.70$ & $< 0.55$
\\
$|g^S_{RL}|$  & $< 0.424$ & $\leq 2$ & $\leq 2$ & $\leq 2$
\\
$|g^S_{LL}|$  & $< 0.55$  & $\leq 2$ & $\leq 2$ & $\leq 2$
\\ \hline
$|g^V_{RR}|$  & $< 0.033$ & $< 0.35$ & $< 0.35$ & $< 0.27$
\\
$|g^V_{LR}|$  & $< 0.060$ & $< 0.29$ & $< 0.23$ & $< 0.20$
\\
$|g^V_{RL}|$  & $< 0.047$ & $< 0.20$ & $< 0.20$ & $< 0.16$
\\
$|g^V_{LL}|$  & $> 0.96$  & $\leq 1$ & $\leq 1$ & $\leq 1$
\\ \hline
\end{tabular}
\end{table}

\begin{table}[bt]
\caption{90\% CL limits  on the $g^V_{\epsilon \omega}$ couplings, 
assuming that  (non-standard) $W$--exchange is the only relevant 
interaction.}
\label{tab:W_couplings}
\centering\vspace{0.2cm}
\begin{tabular}{|l|l|l|l|}
\hline & \hfil $\mu\to e$\hfil &
\hfil $\tau\to\mu$\hfil &\hfil $\tau\to e$ \hfil 
\\\hline
$|g^V_{RR}|$ & $<0.0028$ & $<0.017$ & $<0.011$ 
\\
$|g^V_{LR}|$ & $<0.060$ & $<0.29$ & $<0.23$ 
\\
$|g^V_{RL}|$ & $<0.047$ & $<0.060$ & $<0.047$ 
\\ 
$|g^V_{LL}|$ & $>0.997$ & $>0.95$ & $>0.97$ 
\\ \hline 
\end{tabular}
\end{table}

Table~\ref{tab:coup_CH} assumes that there are no tensor couplings,
i.e. $g^T_{\epsilon\omega}=0$. This condition is satisfied in
any model where the interactions are mediated by vector bosons
and/or charged scalars \cite{PS:95}.
In this case, the quantities  $(1-\frac{4}{3}\rho)$,
$(1-\frac{4}{3}\xi\delta)$ and
$(1-\frac{4}{3}\rho) + \frac{1}{2} (1-\xi)$
reduce to sums of $|g^n_{\epsilon\omega}|^2$,
which are positive semidefinite;
i.e.~,
in the absence of tensor couplings, $\rho\leq\frac{3}{4}$, 
$\xi\delta\leq\frac{3}{4}$
and $(1-\xi) > 2 (\frac{4}{3}\rho - 1)$.

If one only considers $W$--mediated interactions, but admitting the
possibility that the $W$ couples non-universally to leptons of any 
chirality  \cite{PS:95}, the stronger limits in
Table~\ref{tab:W_couplings} are obtained.
In this case, the $g^V_{\epsilon\omega}$ 
constants factorize into the product of two leptonic $W$ couplings, 
implying \cite{MU:85} additional relations among the couplings,
such as $g^V_{LR}\ g^V_{RL} = g^V_{LL}\ g^V_{RR}$, which hold within
any of the three channels, $(\mu, e)$,
$(\tau, e)$, and $(\tau, \mu)$.
Moreover, 
there are additional equations relating different processes,
such as \cite{PS:95}
$g^V_{\mu_L \tau_L}\ g^V_{e_L \tau_R}  =
g^V_{\mu_L \tau_R}\ g^V_{e_L \tau_L}$.
The normalization condition~\eqn{eq:normalization}
provides lower bounds on the $g^V_{LL}$ couplings.

For $W$--mediated interactions,
the hadronic $\tau$--decay modes can also be used to test the
structure of the $\tau\nu_\tau W$
vertex, if one assumes that the W coupling to the light quarks 
is the Standard Model one\footnote{
A more general analysis of the process
$e^+e^-\to\tau^+\tau^-\to (\bar\nu_\tau\pi^+\pi^0 ) (\nu_\tau\pi^-\pi^0 )$,
which includes scalar--like couplings, can be found in 
Ref.~\protect\citenum{TK:93}.
}.
The $\cP_\tau$ dependent part of the decay amplitude is then
proportional to (twice) the mean $\nu_\tau$ helicity
\bel{eq:nu_helicity}
h_{\nu_\tau} \,\equiv\, {|g_R|^2 - |g_L|^2
\over |g_R|^2 + |g_L|^2} ,
\ee
which plays a role analogous to the leptonic--decay parameter $\xi$.
The analysis of $\tau^+\tau^-$ decay correlations
in leptonic--hadronic and hadronic--hadronic decay modes, using
the $\pi$, $\rho$ and $a_1$ hadronic final states, gives
\cite{PDG:96,SLD:97,L3:96,CLEO:97a}
\be
h_{\nu_\tau} = -1.003\pm 0.022 \ .
\ee
This implies $|g_R/g_L|^2 < 0.017$ (90\% CL).

\subsection{Expected Signals in Minimal New--Physics Scenarios}

All experimental results obtained so far are consistent with the
Standard Model. Clearly, the Standard Model
provides the dominant contributions to the $\tau$--decay amplitudes.
Future high--precision measurements of allowed $\tau$--decay modes 
should then look for small deviations of the Standard Model
predictions and
find out the possible source of any detected discrepancy.

In a first analysis, it seems natural to assume \cite{PS:95}
that new physics effects would be dominated by the exchange of a single
intermediate boson, coupling to two leptonic currents.
Table~\ref{tab:summary}
summarizes the expected changes
on the measurable shape parameters \cite{PS:95},
in different new physics scenarios.
The four general cases studied correspond to adding a single intermediate
boson exchange, $V^+$, $S^+$, $V^0$, $S^0$ 
(charged/neutral, vector/scalar), to the Standard Model contribution.

\begin{table}[bht]
\caption{Changes in the Michel parameters induced by
the addition of a single intermediate boson exchange 
($V^+$, $S^+$, $V^0$, $S^0$)
to the Standard Model contribution \protect\cite{PS:95}.}
\label{tab:summary}
\centering\vspace{0.2cm}
\begin{tabular}{|c|c|c|c|c|}
\hline
& $V^+$  &  $S^+$  &  $V^0$  &  $S^0$
\\ \hline
$\rho - 3/4$   & $< 0$  &  0   &  0   &  $< 0$
\\ 
$\xi - 1$      &   $\pm$   & $< 0$ & $< 0$ &  $\pm$ 
\\ 
$\delta\xi-3/4$& $< 0$ & $< 0$ & $< 0$ & $< 0$
\\ 
$\eta$         &   0   &  $\pm$   &   $\pm$  &  $\pm$ 
\\ \hline
\end{tabular}
\end{table}

\section{NEUTRAL--CURRENT COUPLINGS}
\label{sec:nc}\setcounter{equation}{0}

In the Standard Model, tau pair production in $e^+e^-$
annihilation proceeds
through the electromagnetic and weak neutral--current interactions,
\bel{eq:production}
e^+e^-\to\gamma,Z\to \tau^+\tau^- .
\ee
At low energies ($s \ll M_Z^2$),
the production cross--section is only sensitive to the
coupling of the $\tau$ to the photon. From the energy dependence of the
production cross--section near threshold, the spin of the $\tau$ has
been determined \cite{DE:78,DA:78}
to be $1/2$ and its mass has been measured to be \cite{BES:95}
$m_\tau = 1776.96^{+0.18}_{-0.21}{}^{+0.25}_{-0.17}$ MeV.
 
At  high  energies, where the $Z$ contribution is important, the
study  of  the
production cross--section allows to extract information   on
the lepton electroweak parameters.
The $Z$ coupling to the neutral lepton current is given by
\bel{eq:L_nc}
\cL_{\mbox{\rms NC}}^Z \, = \, { g \over 2 \cos{\theta_W}} \,
     Z_\mu \,\sum_l \bar l \gamma^\mu (v_l - a_l \gamma_5) l \, ,
\ee
where 
$v_l = T_3^l (1-4|Q_l|\sin^2{\theta_W})$ and $a_l=T_3^l$;
%
%
i.e., the weak neutral couplings are predicted to be
the same for all leptons with equal electric charge.
%
%

For unpolarized $e^+$ and $e^-$ beams, the differential 
$e^+e^-\to l^+l^-$
cross--section can be written as
\bel{eq:dif_cross}
{d\sigma\over d\Omega}\, = \, {\alpha^2\over 8 s} \, 
         \left\{ A \, (1 + \cos^2{\theta}) \, + B\,  \cos{\theta}\,
     - \, h_l \left[ C \, (1 + \cos^2{\theta}) \, +\, D \cos{\theta}
         \right] \right\} ,
\ee
where $h_l$ ($=\pm1$) is (twice) the $l^-$ helicity and $\theta$ is 
the scattering angle between $e^-$ and $l^-$.
At lowest order,
\be\label{eq:ABCD}\begin{array}{l}
A  =  1 + 2 v_e v_l \,\mbox{\rm Re}(\chi)
 + \left(v_e^2 + a_e^2\right) \left(v_l^2 + a_l^2\right) |\chi|^2, 
\\ 
B  =  4 a_e a_l \,\mbox{\rm Re}(\chi) + 8 v_e a_e v_l a_l  |\chi|^2  , 
\\ 
C  =  2 v_e a_l \,\mbox{\rm Re}(\chi) + 2 \left(v_e^2 + a_e^2\right) 
  v_l a_l |\chi|^2 ,
\\ 
D  =  4 a_e v_l \,\mbox{\rm Re}(\chi) + 4 v_e a_e \left(v_l^2 +
      a_l^2\right) |\chi|^2  , 
\ea\ee
and  $\chi$  contains the $Z$  propagator
\bel{eq:Z_propagator}
\chi \, = \, {G_F M_Z^2 \over 2 \sqrt{2} \pi \alpha }
     \,\, {s \over s - M_Z^2 + i s \Gamma_Z  / M_Z } \, . 
\ee

The coefficients $A$, $B$, $C$ and $D$ can be experimentally determined,
by measuring the total cross--section, the forward--backward asymmetry,
the polarization asymmetry and the forward--backward polarization
asymmetry, respectively:
\beqn\label{eq:sigma}
\sigma(s) & = & 
{4 \pi \alpha^2 \over 3 s } \, A \, ,
\no\\ \label{eq:A_FB}
\cA_{\mbox{\rms FB}}(s)&\equiv & {N_F - N_B \over N_F + N_B}
     =  {3 \over 8} {B \over A}\,  ,
\no\\ \label{eq:A_pol}
\cA_{\mbox{\rms Pol}}(s)  &\equiv &
{\sigma^{(h_l =+1)}
- \sigma^{(h_l =-1)} \over \sigma^{(h_l =+1)} + \sigma^{(h_l = -1)}}
\, = \,  - {C \over A} \, ,
\\ \label{eq:A_FB_pol}
\cA_{\mbox{\rms FB,Pol}}(s) & \equiv & 
{N_F^{(h_l =+1)} - 
N_F^{(h_l = -1)} - N_B^{(h_l =+1)} + N_B^{(h_l = -1)} \over
N_F^{(h_l =+1)} + N_F^{(h_l = -1)} + N_B^{(h_l =+1)} + N_B^{(h_l = -1)}}
\, = \, -{3 \over 8} {D \over A}\, . \quad \no
\eeqn
Here, $N_F$ and $N_B$ denote the number of $l^-$'s
emerging in the forward and backward hemispheres,
respectively, with respect to the electron direction.

For $s = M_Z^2$,
the real part of the $Z$ propagator vanishes
and the photon exchange terms can be neglected
in comparison with the $Z$--exchange contributions
($\Gamma_Z^2 / M_Z^2 \ll 1$). Eqs.~\eqn{eq:sigma}   
become then,
\beqn\label{eq:sigma_Z}
\sigma^{0,l}  \equiv  \sigma(M_Z^2)  = 
 {12 \pi  \over M_Z^2 } \, {\Gamma_e \Gamma_l\over\Gamma_Z^2}\, ,
&& \qquad
\cA_{\mbox{\rms FB}}^{0,l}\equiv\cA_{FB}(M_Z^2) = {3 \over 4}
\cP_e \cP_l \, ,
\no\\ \label{eq:A_pol_Z}
\cA_{\mbox{\rms Pol}}^{0,l} \equiv
  \cA_{\mbox{\rms Pol}}(M_Z^2)  = \cP_l \, ,\quad
&& \qquad
\cA_{\mbox{\rms FB,Pol}}^{0,l} \equiv 
\cA_{\mbox{\rms FB,Pol}}(M_Z^2)  =  {3 \over 4} \cP_e  \, ,\quad\quad
\eeqn
where $\Gamma_l$
is the $Z$ partial decay width to the $l^+l^-$ 
final state, and
\bel{eq:P_l}
\cP_l \, \equiv \, { - 2 v_l a_l \over v_l^2 + a_l^2} 
\ee
is the average longitudinal polarization of the lepton $l^-$,
which only depends on the ratio of the vector and axial--vector couplings.
$\cP_l$ is a sensitive function of $\sin^2{\theta_W}$.

The $Z$ partial decay width to the $l^+l^-$ final state,
\bel{eq:Z_l_QED}
\Gamma_l  \equiv\Gamma(Z\to l^+l^-) = 
{G_F M_Z^3\over 6\pi\sqrt{2}} \, (v_l^2 + a_l^2)\, 
\left(1 + {3\alpha\over 4\pi}\right) ,
\ee
determines the sum $(v_l^2 + a_l^2)$, while the ratio $v_l/a_l$
is derived from the asymmetries\footnote{
The asymmetries determine two possible solutions for $|v_l/a_l|$.
This ambiguity can be solved with lower--energy data or
through the measurement of the transverse
spin--spin correlation \cite{BPR:91}
of the two $\tau$'s in $Z\to\tau^+\tau^-$,
which requires \cite{FSanchez,DELPHI:97} $|v_\tau/a_\tau|<< 1$.}.
The signs of $v_l$ and $a_l$ are fixed by requiring $a_e<0$.

The measurement of the final polarization asymmetries can (only) be done for 
$l=\tau$, because the spin polarization of the $\tau$'s
is reflected in the distorted distribution of their decay products.
Therefore, $\cP_\tau$ and $\cP_e$ can be determined from a
measurement of the spectrum of the final charged particles in the
decay of one $\tau$, or by studying the correlated distributions
between the final products of both $\tau's$ \cite{ABGPR:92}.

With polarized $e^+e^-$ beams, one can also study the left--right
asymmetry between the cross--sections for initial left-- and right--handed
electrons.
At the $Z$ peak, this asymmetry directly measures 
the average initial lepton polarization, $\cP_e$,
without any need for final particle identification:
\bel{eq:A_LR}
\cA_{\mbox{\rms LR}}^0\,\equiv\, \cA_{\mbox{\rms LR}}(M_Z^2)
  \, = \, {\sigma_L(M_Z^2)
- \sigma_R(M_Z^2) \over \sigma_L(M_Z^2) + \sigma_R(M_Z^2)}
\, = \,  - \cP_e \,  .
\ee
%

\begin{table}[tbh]
\centering
\caption{Measured values \protect\cite{LEP:96}
of $\Gamma_l\equiv\Gamma(Z\to l^+l^-)$
and the leptonic forward--backward asymmetries.
The last column shows the combined result 
(for a massless lepton) assuming lepton universality.
\label{tab:LEP_asym}}
\vspace{0.2cm}
\begin{tabular}{|c|c|c|c|c|}
\hline
& $e$ & $\mu$ & $\tau$ & $l$ 
\\ \hline
$\Gamma_l$ \, (MeV) & $83.96\pm 0.15$
& $83.79\pm 0.22$ & $83.72\pm 0.26$ & $83.91\pm 0.11$
\\
$\cA_{\mbox{\rms FB}}^{0,l}$ \, (\%) & $1.60\pm 0.24$
& $1.62\pm 0.13$ & $2.01\pm 0.18$ & $1.74\pm 0.10$
\\ \hline
\end{tabular}
\end{table}
\begin{table}[bth]
\centering
\caption{Measured values \protect\cite{LEP:96}
of the different polarization asymmetries.}
\label{tab:pol_asym}
\vspace{0.2cm}
\begin{tabular}{|c|c|c|c|}
\hline
$\cA_{\mbox{\rms Pol}}^{0,\tau} = \cP_\tau$ &
${4\over 3}\cA^{0,\tau}_{\mbox{\rms FB,Pol}} = \cP_e$ &
$-\cA_{\mbox{\rms LR}}^0 = \cP_e$
& $- \{{4\over 3}\cA_{\mbox{\rms FB}}^{0,l}\}^{1/2} = P_l$
\\ \hline
$-0.1401\pm 0.0067$ & $-0.1382\pm 0.0076$ & $-0.1542\pm 0.0037$
& $-0.1523\pm 0.0044$
\\ \hline
\end{tabular}
\end{table}

\begin{table}[tbh]
\centering
\caption{
Effective vector and axial--vector lepton couplings
derived from LEP and SLD data \protect\cite{LEP:96}.
\label{tab:nc_measured}}
\vspace{0.2cm}
\begin{tabular}{|c|c|c|}     
\hline
& \multicolumn{2}{c|}{Without Lepton Universality}\\ \cline{2-3}
& LEP & LEP + SLD\\ \hline
$v_e$ & $-0.0368\pm 0.0015$ & 
        $-0.03828 \pm 0.00079$
\\
$v_\mu$ & $-0.0372\pm 0.0034$ & 
          $-0.0358 \pm 0.0030$
\\
$v_\tau$ & $-0.0369\pm 0.0016$ & 
           $-0.0367 \pm 0.0016$
\\
$a_e$ & $-0.50130\pm 0.00046$ & 
        $-0.50119 \pm 0.00045$
\\
$a_\mu$ & $-0.50076\pm 0.00069$ & 
          $-0.50086 \pm 0.00068$
\\
$a_\tau$ & $-0.50116\pm 0.00079$ & 
           $-0.50117 \pm 0.00079$
\\ \hline  
$v_\mu/v_e$ & $1.01\pm 0.11$ & 
              $\phantom{-}0.935\pm 0.085$
\\
$v_\tau/v_e$ & $1.001\pm 0.062$ & 
               $\phantom{-}0.959\pm 0.046$
\\
$a_\mu/a_e$ & $0.9989\pm 0.0018$ & 
              $\phantom{-}0.9993\pm 0.0017$
\\
$a_\tau/a_e$ & $0.9997\pm 0.0019$ & 
               $\phantom{-}1.0000\pm 0.0019$
\\ \hline
& \multicolumn{2}{c|}{With Lepton Universality}\\ \cline{2-3}
& LEP & LEP + SLD \\ \hline
$v_l$ & $-0.03688\pm 0.00085$ & 
        $-0.03776 \pm 0.00062$
\\
$a_l$ & $-0.50115\pm 0.00034$ & 
        $-0.50108 \pm 0.00034$
\\
$a_\nu=v_\nu$ & $+ 0.5009\pm 0.0010$ & 
                $+0.5009\pm 0.0010$ 
\\ \hline
\end{tabular}
\end{table}

Tables~\ref{tab:LEP_asym} and \ref{tab:pol_asym}
show the present experimental results
for the leptonic $Z$--decay widths and asymmetries.
The data are in excellent agreement with the Standard Model predictions
and confirm the universality of the leptonic neutral couplings\footnote{
A small 0.2\% difference between $\Gamma_\tau$ and $\Gamma_{e,\mu}$
is generated by the $m_\tau$ corrections.}.
There is however a small ($\sim 2\sigma$) discrepancy between the
$\cP_e$ values obtained \cite{LEP:96} from 
$\cA^{0,\tau}_{\mbox{\rms FB,Pol}}$ 
and $\cA_{\mbox{\rms LR}}^0$.
Assuming lepton universality, 
the combined result from all leptonic asymmetries gives
\bel{eq:average_P_l}
\cP_l = - 0.1500\pm 0.0025 \ .
\ee

The measurement of $\cA_{\mbox{\rms Pol}}^{0,\tau}$ and
$\cA^{0,\tau}_{\mbox{\rms FB,Pol}}$ assumes that the $\tau$ decay
proceeds through the Standard Model charged--current interaction.
A more general analysis should take into account the fact that the
$\tau$ decay width depends on the product $\xi\cP_\tau$, 
where $\xi$
is the corresponding Michel parameter in leptonic decays, or
the equivalent quantity $\xi_h$  ($=-h_{\nu_\tau}$) in the semileptonic 
modes.
A separate measurement of $\xi$ and $\cP_\tau$ has been performed by
ALEPH \cite{ALEPH:94} ($\cP_\tau = -0.139\pm 0.040$)
and L3 \cite{L3:96} ($\cP_\tau = -0.154\pm 0.022$),
using the correlated distribution of the $\tau^+\tau^-$ decays.

\begin{figure}[bht]
\centering
\centerline{\epsfxsize =9.5cm \epsfbox{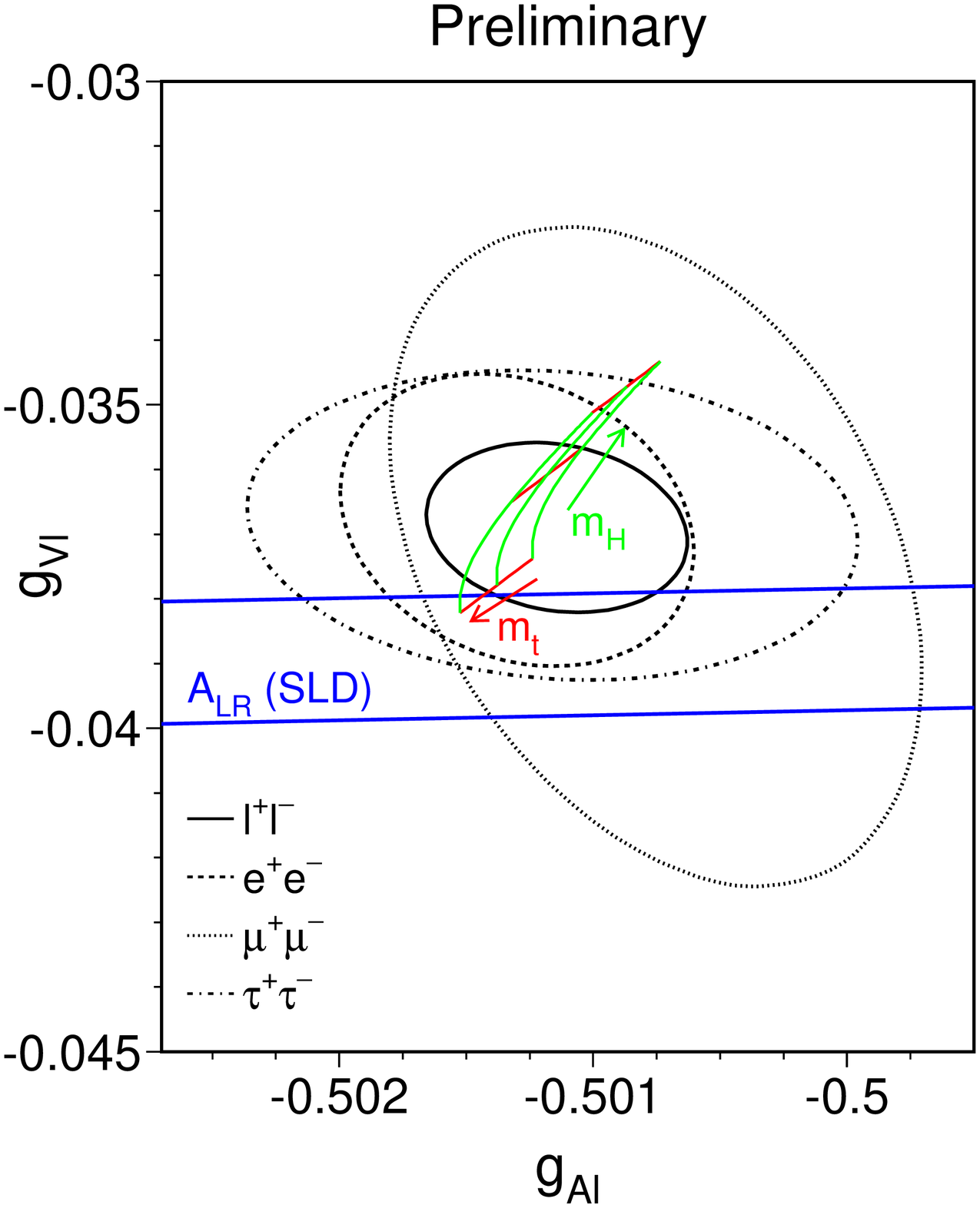}}
\caption{68\% probability contours in the $a_l$-$v_l$ plane
from LEP measurements \protect\cite{LEP:96}. 
The solid contour assumes lepton universality. 
Also shown is the $1\sigma$ band resulting from the
$\protect\cA_{\mbox{\protect\rms LR}}^0$ measurement at SLD. 
The grid corresponds to the Standard Model prediction.} 
\label{fig:gagv}
\end{figure}

The combined analysis of all\footnote{
Not yet included is the
recent SLD measurement \protect\cite{SLD:97b} of the leptonic
forward--backward left--right asymmetries:
$\cP_e = -0.152\pm 0.012\pm 0.001$,
$\cP_\mu = -0.102\pm 0.034\pm 0.002$,
$\cP_\tau = -0.195\pm 0.034\pm 0.003$.
} \
leptonic observables from LEP
and SLD ($\cA_{\mbox{\rms LR}}^0$) results in the
effective vector and axial--vector couplings given \cite{LEP:96} in
Table~\ref{tab:nc_measured}. 
The corresponding 68\% probability contours in the $a_l$--$v_l$ plane 
are shown in Figure~\ref{fig:gagv}.
The measured ratios of the $e$, $\mu$ and $\tau$ couplings
provide a test of charged--lepton universality in the neutral--current 
sector.

The neutrino couplings can be determined from the invisible 
$Z$ decay width, by assuming three identical neutrino generations
with left--handed couplings (i.e., $v_\nu=a_\nu$), 
and fixing the sign from neutrino scattering 
data \cite{CHARMII:94}.
The resulting experimental value \cite{LEP:96},
given in Table~\ref{tab:nc_measured},
is in perfect agreement with the Standard Model.
Alternatively, one can use the Standard Model prediction for 
$\Gamma_{\mbox{\rms inv}}/\Gamma_l$
to get a determination of the number of (light) neutrino flavours
\cite{LEP:96}:
\be
N_\nu = 2.989\pm 0.012 \, .
\ee
The universality of the neutrino couplings has been tested
with $\nu_\mu e$ scattering data, which fixes \cite{CHARMII:94b}
the $\nu_\mu$ coupling to the $Z$: \ 
$v_{\nu_\mu} =  a_{\nu_\mu} = 0.502\pm 0.017$.

The measured leptonic asymmetries can be used to obtain the
effective electroweak mixing angle in the charged--lepton sector: 
\cite{LEP:96}
\bel{eq:bar_s_W_l}
\sin^2{\theta^{\mbox{\rms lept}}_{\mbox{\rms eff}}} \equiv
{1\over 4}   \left( 1 - {v_l\over a_l}\right)  = 0.23114\pm 0.00031 \, .
\ee
Including also the hadronic asymmetries, one gets \cite{LEP:96}
$\sin^2{\theta^{\mbox{\rms lept}}_{\mbox{\rms eff}}} =
0.23165\pm 0.00024$ 
with a $\chi^2/\mbox{\rm d.o.f.} = 12.8/6$.

\goodbreak 
\section{ELECTROMAGNETIC AND WEAK MOMENTS}
\label{sec:structure}\setcounter{equation}{0}

A general description of the electromagnetic coupling of a 
spin--$\frac{1}{2}$ charged lepton to the virtual photon
involves three different form factors:
\bel{eq:em_ff}
T[l\bar l \gamma^*] = e \, \varepsilon_\mu(q) \, \bar l
\left[F_1(q^2)\gamma^\mu
+ i{F_2(q^2)\over 2 m_l} \sigma^{\mu\nu}q_\nu +
{F_3(q^2)\over 2 m_l} \sigma^{\mu\nu}\gamma_5 q_\nu\right] l \ ,
\ee
where $q^\mu$ is the photon momentum.
Owing to the conservation of the electric charge,
$F_1(0)=~1$.
At $q^2=0$, the other two form factors reduce to the
lepton magnetic dipole moment, 
$\mu_l\equiv (e /2 m_l) \, (g^\gamma_l/2) = e (1+F_2(0))/2 m_l$,
and electric dipole moment $d_l^\gamma = e F_3(0)/2 m_l$.
Similar expressions can be defined for the $l\bar l$
coupling to a virtual $Z$.

The $F_i(q^2)$ form factors are sensitive quantities to a
possible lepton substructure.
Moreover, $F_3(q^2)$ violates $T$ and $P$ invariance; 
thus, the electroweak dipole moments
$d^{\gamma,Z}_\tau$, which vanish in the Standard Model, constitute
a good probe of CP violation.
Owing to their chiral--changing structure, the dipole moments 
may provide important insights on the mechanism responsible for
mass generation. In general, one expects \cite{MA:94}
that a fermion of mass
$m_f$ (generated by physics at some scale $M\gg m_f$) will have
induced dipole moments proportional to some power of $m_f/M$.
Therefore, heavy fermions such as the $\tau$ should be a good 
testing ground for this kind of effects.

Information on the $\tau$ electroweak form factors can be obtained
by measuring the $e^+ e^- \to \tau^+ \tau^-$ cross--section.
Their $q^2=0$ values can be tested in 
$e^+ e^- \to \tau^+\tau^-\gamma$ and in the decay
$Z\to\tau^+ \tau^-\gamma$ \cite{GM:91}.
A general analysis of the $\tau$ electroweak 
form factors has never been performed.
The existing experimental tests only provide limits on a single
$F_i$ assuming the other form factors to take their
Standard Model values.

At low energies, where the $Z$ contribution is very small,
the deviations from the QED prediction are usually parametrized
through
$F_1(s) = [1\mp s/(s-\Lambda^2_\pm)]$.
The cut-off parameters  $\Lambda_{\pm}$
characterize the validity of QED and measure the
point-like nature of the $\tau$.
From PEP and PETRA data, one finds  \cite{MA:90} 
$\Lambda_+(\tau)> 285$ GeV and 
$\Lambda_-(\tau)> 246$ GeV (95\% CL),
which correspond to upper limits on the $\tau$ charge
radius of  $10^{-3}$ fm.

The same PEP/PETRA data can be used to extract limits on
the $\tau$ anomalous magnetic moment  \cite{SI:83,MA:89},
$a^\gamma_\tau \equiv (g^\gamma_\tau - 2)/2$,
or electric dipole moment \cite{AS:90}; one finds:
$|a^\gamma_\tau|  < 0.023$  (95\% CL), 
$|d^\gamma_\tau| < 1.6 \times 10^{-16} \,  e$ cm 
(90\% CL).           
These limits actually probe the corresponding form factors $F_2(s)$
and $F_3(s)$ at $s\sim 35$ GeV.
More direct bounds at $q^2=0$ have been extracted \cite{WE:96}
from the decay\footnote{
The present upper limit on $|a_\tau^\gamma|$ has been extracted from the
dependence of $\Gamma(Z\to\tau^+\tau^-\gamma)$ on $|a_\tau^\gamma|^2$,
neglecting the interference terms, which are linear in $|a_\tau^\gamma|$
but are suppressed by a factor $m_\tau^2/M_Z^2$.
This approximation is no longer justified if the limit is better than
a few per cent \protect\cite{BiRi:96}.} \
 $\, Z\to\tau^+\tau^-\gamma$:
\be
|a_\tau^\gamma| < 0.0104  \, , \qquad\qquad
|d^\gamma_\tau| < 5.8 \times 10^{-17} \,  e \,\mbox{\rm cm} 
\qquad\quad (95\%\,\mbox{\rm CL}) \, .
\ee
Slightly better, but more model--dependent, limits have been derived
\cite{EM:93} from the $Z\to\tau^+\tau^-$ decay width:
$-0.004 < a_\tau^\gamma < 0.006$, 
$|d^\gamma_\tau| < 2.7 \times 10^{-17} \,  e \, \mbox{\rm cm}$
(95\% CL);
these bounds would be invalidated in the presence of any CP--conserving
contribution to $\Gamma(Z\to\tau^+\tau^-)$ interfering destructively
with the Standard Model amplitude.

In the Standard Model, $d^\gamma_\tau$ vanishes, while
the overall value of $a^\gamma_\tau$
is dominated by the  second order QED contribution \cite{SC:48},
$a^\gamma_\tau \approx \alpha / 2 \pi$.
Including QED corrections up to O($\alpha^3$),
hadronic vacuum polarization contributions
and the corrections due to the weak interactions 
(which are a factor 380
larger than for the muon), the $\tau$ anomalous magnetic moment has been
estimated to be \cite{NA:78,SLM:91}
\bel{eq:a_th_tau}
a^\gamma_\tau\big |_{th} \, = \, (1.1773 \pm 0.0003)
     \times 10^{-3} \, .
\ee

The first direct limit on the weak anomalous magnetic moment
has been obtained by L3, by using correlated azimuthal
asymmetries of the $\tau^+\tau^-$ decay products \cite{BGV:94}. 
The
preliminary (95\% CL) result of this analysis is \cite{ESanchez}:
\bel{eq:a_Z}
 -0.016 < a^Z_\tau < 0.011 \ .
\ee

The possibility of a CP--violating weak dipole moment of the $\tau$ has
been investigated at LEP, by studying
$T$--odd triple correlations \cite{BN:89,BE:91}
of the final $\tau$--decay products in $Z\to\tau^+\tau^-$ events.
The present (95\% CL) limits are \cite{WE:96}:
\bel{eq:d_Z_tau}
\ba
\vert\mbox{\rm Re}\, d^Z_\tau(M_Z^2)\vert \le 3.6\times 10^{-18}
   \, e \,  \mbox{\rm cm} \, ,
\\ 
\vert\mbox{\rm Im}\, d^Z_\tau(M_Z^2)\vert \le 1.1\times 10^{-17}
   \, e \,  \mbox{\rm cm} \, .
\ea\ee
These limits provide useful constraints on different models
of CP violation \cite{BN:89,BLMN:89,DR:91,KOR:91}.

T--odd signals can be also generated through
a relative phase between the vector and axial--vector couplings
of the $Z$ to the $\tau^+\tau^-$ pair \cite{BPR:91}, i.e.
$\mbox{\rm Im}(v_\tau a_\tau^*) \not= 0$.
This effect, which in the  Standard Model appears \cite{BR:89} at the
one--loop level through absorptive parts in the electroweak amplitudes,
gives rise \cite{BPR:91} to a spin--spin correlation associated with the
transverse (within the production plane) and normal (to the production
plane) polarization components of the two $\tau$'s.
A preliminary analysis of this transverse--normal spin correlation has been
reported by  ALEPH \cite{FSanchez}.

\section{CP VIOLATION}
\label{sec:cp}\setcounter{equation}{0}

In the three--generation Standard Model, the violation of the CP symmetry 
originates from the single phase 
naturally occurring in the quark mixing matrix \cite{KM:73} . 
Therefore, CP violation
is predicted to be absent in the lepton sector (for massless neutrinos).
The present experimental observations are in agreement with the
Standard Model; nevertheless, the correctness of the
Kobayashi---Maskawa mechanism is far from being proved.
Like fermion masses and quark mixing angles, the origin of the
Kobayashi---Maskawa phase lies in the most obscure part of the
Standard Model Lagrangian: the scalar sector.
Obviously, CP violation could well be a sensitive probe for new
physics.

Up to now, CP violation in the lepton sector
has been investigated mainly through the electroweak
dipole moments.
Violations of the CP symmetry could also happen in the
$\tau$ decay amplitude.
In fact, the possible CP--violating effects can be expected to be larger 
in $\tau$ decay than in $\tau^+\tau^-$ production \cite{tsai}. 
Since the decay of the $\tau$ proceeds through a weak interaction,
these effects could be ${\cal O}(1)$ or ${\cal O}(10^{-3})$,
if the leptonic CP violation is {\it weak} or {\it milliweak}
 \cite{tsai}.

With polarized electron (and/or positron) beams, one could use the
longitudinal polarization vectors of the incident leptons to construct
T--odd rotationally invariant products. CP could be tested by comparing
these T--odd products in $\tau^-$ and $\tau^+$ decays.
In the absence of beam polarization, CP violation could still be tested
through $\tau^+\tau^-$ correlations. In order to separate possible 
CP--odd effects in the $\tau^+\tau^-$ production and in the $\tau$ decay,
it has been suggested to study the final decays of the $\tau$--decay
products and build the so-called 
{\it stage--two spin--correlation functions} \cite{stscf}.
For instance, one could study the chain process
$e^+e^-\to\tau^+\tau^-\to (\rho^+\bar\nu_\tau)(\rho^-\nu_\tau)\to
(\pi^+\pi^0\bar\nu_\tau)(\pi^-\pi^0\nu_\tau)$. The distribution of the
final pions provides information on the $\rho$ polarization, which allows
to test for possible CP--violating effects in the 
$\tau\to\rho\nu_\tau$ decay.

CP violation could also be tested
through rate asymmetries, i.e. comparing the partial fractions
$\Gamma(\tau^-\to X^-)$ and $\Gamma(\tau^+\to X^+)$. However, this kind
of signal requires the presence of strong final--state interactions in the
decay amplitude.
Another possibility would be to study T--odd (CPT--even) asymmetries in the
angular distributions of the final hadrons in semileptonic 
$\tau$ decays \cite{FM:96}. 
Explicit studies of the decay modes \cite{KKSW:94}
$\tau^-\to K^-\pi^-\pi^+\nu_\tau, \pi^- K^- K^+\nu_\tau$ and \cite{ChHT:95}
$\tau^-\to \pi^-\pi^-\pi^+\nu_\tau$  show that
sizeable CP--violating effects
could be generated in some models of CP violation 
involving several Higgs doublets
or left--right symmetry.

\section{LEPTON--NUMBER VIOLATION}
\label{sec:L-violation}\setcounter{equation}{0}

In the minimal Standard Model with massless neutrinos, there is a
separately conserved additive lepton number for each generation. All
present data are
consistent with this conservation law. However, there are no strong
theoretical
reasons forbidding a mixing among the  different  leptons, in the same
way as happens in the quark sector.
Many models in fact predict lepton--flavour or even
lepton--number violation at some level
\cite{MO:92,GV:92,IKP:95,ChR:96,KKL:97}.
Experimental searches for these processes
can provide information on the scale at which the new physics begins to
play a  significant role.

$K$, $\pi$ and $\mu$ decays, together with $\mu$--$e$ conversion,
neutrinoless double beta
decays  and  neutrino  oscillation  studies,  have  put  already
stringent  limits \cite{PDG:96} on
lepton--flavour and lepton--number violating interactions.
However, given the present lack of
understanding of the origin of fermion generations, one can imagine
different
patterns of violation of this conservation law  for  different  mass
scales.
Moreover, the larger mass of the $\tau$ opens the possibility of 
new types of decay which are kinematically forbidden for the $\mu$.

\begin{table}[thb]
\caption{Present limits 
\protect\cite{PDG:96,CLEO:97b}
on the branching ratios of lepton--flavour and lepton--number violating
decays of the $\tau$.
All bounds are at $90\% $ CL, except the $l^-G^0$ modes which are at
$95\% $ CL ($G^0$ denotes an unobservable neutral particle).} 
\label{tab:LNV}
\centering\vspace{0.2cm}
\begin{tabular}{|l|r||l|r|}
\hline
Decay mode & Upper limit & Decay mode & Upper limit \\ \hline
$\tau^-\to e^-e^+e^-$     & $3.3 \times 10^{-6}$ &
$\tau^-\to\mu^-\mu^+\mu^-$& $1.9 \times 10^{-6}$  \\
$\tau^-\to e^-\mu^+\mu^-$ & $3.6 \times 10^{-6}$  &
$\tau^-\to \mu^- e^+ e^-$ & $3.4 \times 10^{-6}$ \\
$\tau^-\to e^+\mu^-\mu^-$ & $3.5 \times 10^{-6}$  &
$\tau^-\to \mu^+ e^- e^-$ & $3.4 \times 10^{-6}$ \\
$\tau^-\to e^-\pi^+\pi^-$ & $4.4 \times 10^{-6}$ &
$\tau^-\to\mu^-\pi^+\pi^-$& $7.4 \times 10^{-6}$ \\
$\tau^-\to e^+\pi^-\pi^-$ & $4.4 \times 10^{-6}$ &
$\tau^-\to\mu^+\pi^-\pi^-$& $6.9 \times 10^{-6}$ \\
$\tau^-\to e^-\rho^0$     & $4.2 \times 10^{-6}$ &
$\tau^-\to \mu^-\rho^0$   & $5.7 \times 10^{-6}$ \\
$\tau^-\to e^-\pi^+ K^-$  & $7.7 \times 10^{-6}$  &
$\tau^-\to\mu^-\pi^+ K^-$ & $8.7 \times 10^{-6}$ \\
$\tau^-\to e^-\pi^- K^+$  & $4.6 \times 10^{-6}$  &
$\tau^-\to\mu^-\pi^- K^+$ & $1.5 \times 10^{-5}$ \\
$\tau^-\to e^+\pi^- K^-$  & $4.5 \times 10^{-6}$  &
$\tau^-\to\mu^+\pi^- K^-$ & $2.0 \times 10^{-5}$ \\
$\tau^-\to e^- K^{*0}$    & $6.3 \times 10^{-6}$ &
$\tau^-\to \mu^- K^{*0}$  & $9.4 \times 10^{-6}$  \\
$\tau^-\to e^- \bar K^{*0}$    & $1.1 \times 10^{-5}$ &
$\tau^-\to \mu^- \bar K^{*0}$  & $8.7 \times 10^{-6}$  \\
$\tau^-\to e^- K^0$       & $1.3 \times 10^{-3}$ &
$\tau^-\to \mu^- K^0$     & $1.0 \times 10^{-3}$  \\
$\tau^-\to e^-\gamma$     & $2.7 \times 10^{-6}$ &
$\tau^-\to \mu^-\gamma$   & $3.0 \times 10^{-6}$ \\
$\tau^-\to e^-\pi^0$      & $3.7 \times 10^{-6}$ &
$\tau^-\to \mu^-\pi^0$    & $4.0 \times 10^{-6}$  \\
$\tau^-\to e^-\eta$       & $8.2 \times 10^{-6}$ &
$\tau^-\to \mu^-\eta$     & $9.6 \times 10^{-6}$ \\
$\tau^-\to e^-\pi^0\pi^0$      & $6.5 \times 10^{-6}$ &
$\tau^-\to \mu^-\pi^0\pi^0$    & $1.4 \times 10^{-5}$  \\
$\tau^-\to e^-\eta\eta$       & $3.5 \times 10^{-5}$ &
$\tau^-\to \mu^-\eta\eta$     & $6.0 \times 10^{-5}$ \\
$\tau^-\to e^-\pi^0\eta$       & $2.4 \times 10^{-5}$ &
$\tau^-\to \mu^-\pi^0\eta$     & $2.2 \times 10^{-5}$ \\
$\tau^-\to e^- G^0$       & $2.7 \times 10^{-3}$ &
$\tau^-\to \mu^- G^0$     & $5 \times 10^{-3}$ \\
$\tau^-\to \bar p \gamma$ & $2.9 \times 10^{-4}$ &
$\tau^-\to \bar p \pi^0$  & $6.6 \times 10^{-4}$ \\
$\tau^-\to \bar p \eta$  & $1.3 \times 10^{-3}$ &&\\
%
\hline
\end{tabular}
\end{table}

The present upper limits on lepton--flavour and
lepton--number violating decays of the $\tau$ are given in 
Table~\ref{tab:LNV}.
These limits are in the range of $10^{-4}$ to $10^{-6}$, which is far
away from the impressive (90\% CL)
bounds \cite{PDG:96} obtained in $\mu$ decay:
\be\begin{array}{l}
\mbox{\rm Br}(\mu^-\to e^- \gamma)  < 4.9 \times 10^{-11}\, , \\
\mbox{\rm Br}(\mu^-\to e^- e^+ e^-) < 1.0 \times 10^{-12}\, , \\
\mbox{\rm Br}(\mu^-\to e^-\gamma\gamma) <  7.2 \times 10^{-11} \, .
\ea\ee
With future $\tau$--decay samples of $10^7$ events
per year, an improvement of one to two orders of magnitude seems possible.

The lepton--flavour violating couplings of the
$Z$ boson can be investigated at LEP. The present ($95\% $ CL)
limits are \cite{OPAL:95c}
\beqn\label{eq:Z_LFV}
\mbox{\rm Br}(Z\to e^\pm\mu^\mp) &<& 1.7 \times 10^{-6} ; \no\\
\mbox{\rm Br}(Z\to e^\pm\tau^\mp) &<& 9.8 \times 10^{-6}  ; \\
\mbox{\rm Br}(Z\to \mu^\pm\tau^\mp) &<& 1.7 \times 10^{-5}  .\no
\eeqn
Below the $Z$ pole, the search for the lepton--flavour violating 
processes $e^+e^-\to e^+\tau^-$ and $e^+e^-\to\mu^+\tau^-$ 
has given the ($95\% $ CL) upper bounds \cite{MARKII:91}:
\bel{eq:LFV_g}\ba
\sigma_{e\tau} / \sigma_{\mu\mu}
< 1.8 \times 10^{-3}  \, ; \\ 
    \sigma_{\mu\tau} / \sigma_{\mu\mu} < 6.1 \times 10^{-3} \, .
\ea\ee

\goodbreak
\section{THE TAU NEUTRINO}
\label{sec:neutrino}\setcounter{equation}{0}

All observed $\tau$ decays are supposed to be accompanied by neutrino
emission, in order to fulfil energy--momentum conservation requirements.
As seen in Sections~\ref{sec:current} and \ref{sec:nc}, the present data
are consistent with the $\nu_\tau$
being a conventional sequential neutrino. Since taus are not produced
by $\nu_e$ or $\nu_\mu$ beams, we know that $\nu_\tau$
is different from the electronic and  muonic
neutrinos, and an upper limit can be set on the couplings of the tau to
$\nu_e$ and $\nu_\mu$: \cite{E531:86}
\bel{eq:taunulimits}
|g_{\tau\nu_e}| < 0.073 \ , \qquad |g_{\tau\nu_\mu}| <
        0.002 \ , \qquad (90\% \,\mbox{\rm CL}).
\ee
These limits can be interpreted in terms of $\nu_e/\nu_\mu\to\nu_\tau$
oscillations, to exclude a
region in the neutrino mass--difference and neutrino mixing--angle space.
In the extreme situations of large $\delta m^2$ or maximal mixing, the
present limits are \cite{E531:86}
\beqn\label{eq:numu_nutau}
\nu_\mu\to\nu_\tau : \qquad
\sin^2{2\theta_{\mu,\tau}} &\! <&\! 0.004 \qquad\; 
    (\mbox{\rm large}\, \delta m_{\mu,\tau}^2) , 
\no\\ 
\delta m_{\mu,\tau}^2 &\! <&\!  0.9 \,\mbox{\rm eV}^2 \quad\;\; 
   (\sin^2{2\theta_{\mu,\tau}} = 1) ;
\\ \label{eq:nue_nutau}
\nu_e\to\nu_\tau : \qquad
\sin^2{2\theta_{e,\tau}} &\! <&\!  0.12 \qquad\;\;\; 
    (\mbox{\rm large}\, \delta m_{e,\tau}^2) , 
\no \\ 
\delta m_{e,\tau}^2 &\! <&\!  9 \,\mbox{\rm eV}^2 \qquad\; 
   (\sin^2{2\theta_{e,\tau}} = 1) .
\eeqn
The new CHORUS \cite{CHORUS} and NOMAD \cite{NOMAD} experiments,
presently running at CERN, and the future Fermilab E803 experiment 
\cite{E803}
are expected to improve the
$\nu_\mu\to\nu_\tau$  oscillation limits 
by at least an order of magnitude.

LEP and SLC have confirmed \cite{LEP:96}
the existence of three (and only
three) different light neutrinos, with standard couplings to the $Z$
(see Section~\ref{sec:nc}).
However,
no direct observation of $\nu_\tau$, that is, interactions resulting
from neutrinos produced in $\tau$ decay, has been made so far.

The expected source of $\tau$ neutrinos in beam dump experiments is
the decay of $D_s$ mesons produced by interactions in the dump; i.e.,
$p+N\to D_s + \cdots $,
followed by the decays $D_s^-\to\tau^-\bar\nu_\tau$
and $\tau^-\to\nu_\tau + \cdots $\  
Several experiments \cite{BEBC:87}
have searched for
$\, \nu_\tau + N \to \tau^- + \cdots $\
interactions with negative results; therefore, only an upper
limit on the production of $\nu_\tau$'s has been obtained.
The direct detection of the $\nu_\tau$
should be possible \cite{dRR:84} at the LHC,
thanks to the large charm--production cross--section of this collider.

The possibility of a non-zero neutrino mass is obviously a
very important
question in particle physics. There is no fundamental principle requiring
a null mass for the neutrino. On the contrary, many extensions of the 
Standard Model
predict non-vanishing neutrino masses, which could have, in addition,
important implications in cosmology and astrophysics.
 
The first attempts to place a limit on $m_{\nu_\tau}$
were done by studying the endpoint of the
momentum spectrum of charged leptons from the decays
$\tau^-\to\nu_\tau l^- \bar\nu_l \quad (l = e,\mu)$. The
precision
which can be achieved is limited by the experimental momentum resolution
of fastest particles, which deteriorates with increasing
centre--of--mass energy.
Better limits have been set by studying the endpoint of the hadronic
mass spectrum of high multiplicity $\tau$ decays.
The limiting factor is then the resolution of the effective
hadronic--mass determination. 
The strongest bound up to date is the preliminary ALEPH limit
\cite{Passalacqua}, 
\be\label{eq:numasslimit}
m_{\nu_\tau} \, < \, 18.2 \,\mbox{\rm MeV} \quad (95\%\,
\mbox{\rm CL}),
\ee
obtained from a two--dimensional likelihood fit of the
visible energy and the invariant--mass distribution of
$\tau^-\to (3 \pi)^-\nu_\tau, (5 \pi)^-\nu_\tau$ 
events. 
 
For comparison, the present limits on the muon and electron
neutrinos are \cite{PDG:96}
$m_{\nu_\mu} < 170$ KeV    (90\% C.L.)
and $m_{\nu_e} < 15$ eV.   
Note, however, that in many models a mass hierarchy among
different generations is expected, with the neutrino mass being
proportional to some power of the mass of its charged lepton partner.
Assuming for instance the  fashionable  relation
$\, m_{\nu_\tau} / m_{\nu_e} \sim (m_\tau/m_e)^2$,
the bound \eqn{eq:numasslimit} would be equivalent to
a limit of 1.5 eV for $m_{\nu_e}$.
A relatively crude measurement of $m_{\nu_\tau}$
may then imply strong constraints on neutrino--mass model building.

More stringent (but model--dependent) bounds on $m_{\nu_\tau}$ can be
obtained from cosmological considerations.
A stable neutrino (or an unstable neutrino with a lifetime comparable
to or longer than the age of the Universe) must not overclose the
Universe. Therefore, measurements of the age of the
Universe exclude stable neutrinos in  the range\cite{GZ:72,LW:77}
200 eV $< m_\nu <$ 2 GeV.
Unstable neutrinos with lifetimes longer than 300 sec
could increase the expansion rate of the Universe, 
spoiling the successful predictions for the
primordial nucleosynthesis of light isotopes in the early universe 
\cite{KO:91}; the mass range
0.5 MeV $< m_{\nu_\tau} < $ 30 MeV
has been excluded in that case \cite{KO:91,DR:93,KA:94,DGT:94,GT:95}. 
For neutrinos of any lifetime decaying into electromagnetic daughter
products, it is possible to exclude the same mass range, combining the
nucleosynthesis constraints with limits based on the supernova SN 1987A
and on BEBC data \cite{DGT:94,GT:95}.
Light neutrinos ($m_{\nu_\tau}< 100$ keV)
decaying through $\nu_\tau\to\nu_\mu + G^0$, 
are also excluded by the nucleosynthesis constraints, if their lifetime
is shorter than $10^{-2}$ sec \cite{KA:94}.

The astrophysical and cosmological arguments lead indeed to quite
stringent limits; however, they always involve (plausible) assumptions
which could be relaxed in some physical scenarios 
\cite{MS:95,FKO:96,HM:96}.
For instance, in deriving the abundance of massive $\nu_\tau$'s at 
nucleosynthesis, it is always assumed that $\tau$ neutrinos annihilate at
the rate predicted by the Standard Model.
Moreover, the present observational situation is rather unclear, due
to the existence of inconsistent sets of data on the primordial
abundances of light isotopes \cite{RRW:97}; therefore, one cannot be
confident in the reliability of such limits.

A $\nu_\tau$ mass in the few MeV range (i.e. the mass sensitivity
which can be achieved in the foreseeable future) could have a host
of interesting astrophysical and cosmological consequences \cite{GT:95}:
relaxing the big-bang nucleosynthesis bound to the baryon density and
the number of neutrino species; allowing big-bang nucleosynthesis to
accommodate a low ($< 20\% $) ${}^4$He mass fraction or high ($>10^{-4}$)
deuterium abundance; improving significantly the agreement between the
cold dark matter theory of structure formation and observations
\cite{DGT:94b};
and helping to explain how type II supernovae explode.

  The electromagnetic structure of the $\nu_\tau$ can be tested through
the process $e^+e^-\to\nu_\tau\bar\nu_\tau\gamma$. The combined data
from PEP and PETRA implies \cite{GR:88} the following $90 \% $ CL upper
bounds on the magnetic moment and charge radius  of the $\nu_\tau$
($\mu_B \equiv e\hbar / 2 m_e$):
$|\mu(\nu_\tau)|  <  4 \times 10^{-6} \, \mu_B$;
$<r^2>(\nu_\tau) <  2 \times 10^{-31} \, \mbox{\rm cm}^2$.
A better limit on the $\nu_\tau$ magnetic moment,
\bel{eq:mu_nu_tau}
|\mu(\nu_\tau)|  <  5.4 \times 10^{-7} \, \mu_B \quad (90\%\,\mbox{\rm CL}) ,
\ee
has been placed by the BEBC experiment \cite{BEBC:92},
by searching for elastic $\nu_\tau e$ scattering events, using a
neutrino beam from a beam dump which has a small $\nu_\tau$ component.

A big $\nu_\tau$ magnetic moment of about $10^{-6} \mu_B$ has been
suggested, in order to make the $\tau$ neutrino an acceptable cold
dark matter candidate. For this to be the case, hovewer, the
$\nu_\tau$ mass should be in the range
1 MeV $ < m_{\nu_\tau} < $ 35 MeV  \cite{GI:90}.
The same region of $m_{\nu_\tau}$ has been suggested in trying to
understand the baryon--antibaryon asymmetry of the universe \cite{CKN:91}.

\setcounter{equation}{0}
\section{HADRONIC DECAYS}
\label{sec:hadronic}

The $\tau$ is the only presently known lepton massive enough to decay
into hadrons. Its semileptonic decays are then an ideal laboratory
for studying the hadronic weak currents in very clean conditions.
The decay modes $\tau^-\to\nu_\tau H^-$
probe the  matrix
element of the left--handed charged current between the vacuum and the
final hadronic state $H^-$,
\bel{eq:Had_matrix}
\langle H^-| \bar d_\theta \gamma^\mu (1-\gamma_5) u | 0 \rangle\, .
\ee
Contrary to the well--known process\ $e^+e^-\to\gamma\to$ hadrons,
which only tests the electromagnetic vector current, the semileptonic
$\tau$--decay modes offer the possibility to study the properties of both
vector and axial--vector currents.
 
For the decay modes with lowest multiplicity,
$\tau^-\to\nu_\tau\pi^-$  and $\tau^-\to\nu_\tau K^-$, the  relevant
matrix  elements  are  already  known  from  the  measured  decays
$\pi^-\to\mu^-\bar\nu_\mu$  and  $K^-\to\mu^-\bar\nu_\mu$,
\bel{eq:f_pi_K}\begin{array}{ccc}
\langle\pi^-(p)|\bar d \gamma^\mu\gamma_5 u | 0 \rangle &
\equiv & -i \sqrt{2} f_\pi p^\mu \, , \\
\langle K^-(p)|\bar s \gamma^\mu\gamma_5 u | 0 \rangle &
\equiv & -i \sqrt{2} f_K p^\mu \, . 
\ea\ee
The corresponding $\tau$ decay widths can then be predicted
rather accurately [Eqs.~\eqn{eq:R_tp} and \eqn{eq:R_tk}].
As shown in Table~\ref{tab:univtm}, these predictions are in good 
agreement with the measured values, and provide a quite precise test
of charged--current universality.

Alternatively, the measured ratio between the
$\tau^-\to\nu_\tau K^-$ and $\tau^-\to\nu_\tau \pi^-$
decay widths can be used to obtain a value for
$\tan^2{\theta_C}\, (f_K / f_\pi)^2$:
\bel{eq:theta_C}
\left|{V_{us}\over V_{ud}}\right|^2
 \, \left( {f_K \over f_\pi} \right)^2 \, =
    \, \left( {m^2_\tau - m^2_\pi \over m^2_\tau - m^2_K } \right)^2
    \, { \Gamma (\tau^-\to\nu_\tau K^- ) \over \Gamma (\tau^-\to\nu_\tau
    \pi^- ) } 
\, {1+ \delta R_{\tau/\pi}\over 1+ \delta R_{\tau/K}}
\, = \, (7.2 \pm 0.3) \times 10^{-2} \, . 
\ee
This number is consistent with (but less precise than) the result
$(7.67 \pm 0.06) \times 10^{-2}$  obtained from  \cite{PDG:96}
$\Gamma (K^- \to \mu^-\bar\nu_\mu ) / \Gamma (\pi^-\to\mu^-\bar\nu_\mu )$.
 
For the Cabibbo--allowed modes with $J^P = 1^-$, the matrix element of
the vector charged current can also be obtained, 
through an isospin  rotation, from the
isovector  part  of  the $e^+ e^-$ annihilation  cross--section  into
hadrons, which
measures the hadronic matrix element of   the  $I=1$  component
of the electromagnetic current,
\bel{eq:em_matrix}
\langle H^0|(\bar u \gamma^\mu u - \bar d \gamma^\mu d)|0\rangle\, .
\ee
The $\tau\to \nu_\tau V^-$ decay width is then expressed as an integral over
the corresponding $e^+ e^-$ cross-section \cite{TS:71,ThS:71}:
%
\bel{eq:cvc}
R_{\tau\to V}  \equiv
{\Gamma (\tau^-\to\nu_\tau V^-) \over \Gamma_{\tau\to e}}
=  {3 \cos^2{\theta_C} \over 2 \pi \alpha^2 m_\tau^8 }\, S_{EW}\,
   \int_0^{m_\tau^2} \, ds \, (m_\tau^2 - s)^2 (m_\tau^2 + 2 s) \, s \,
    \sigma^{I=1}_{e^+ e^- \to V^0}(s) \, , 
\ee
where the factor $S_{EW}=1.0194$ contains the renormalization--group 
improved
electroweak correction at the leading logarithm approximation \cite{MS:88}.
Using the available \ $e^+ e^- \to$ hadrons \ data,
one can then predict the $\tau$ decay widths for these modes
\cite{GRM:85,KS:90,EI:95,NP:93,EI:96}.

\begin{table}[htb]
\centering
\caption{$R_{\tau\to V}$
from $\tau$--decay \protect\cite{CO:96,PDG:96} and $e^+e^-$ data
\protect\cite{EI:96}.}
\label{tab:vector}
\vspace{0.2cm}
\begin{tabular}{|c|c|c|}
\hline
$V^-$   & \multicolumn{2}{c|}{
$R_{\tau\to V}\equiv \Gamma (\tau^-\to\nu_\tau V^-) / \Gamma_{\tau\to e}$}
\\
\cline{2-3}    &  $\tau^-\to\nu_\tau V^-$ & $e^+e^-\to V^0$ 
\\ \hline 
$\pi^-\pi^0$  & $1.413\pm0.012$ & $1.360\pm0.043$ 
\\
$2\pi^-\pi^+\pi^0$ & $0.239\pm0.005$ & $0.239\pm0.031$ 
\\
$\pi^-3\pi^0$ & $0.064\pm0.008$ & $0.059\pm0.006$ 
\\
$\pi^-\omega$ & $0.108\pm0.004$ & $0.098\pm0.011$ 
\\
$3\pi^-2\pi^+\pi^0$ &	$0.0012\pm0.0003$ & $\geq 0.0010$
\\
$(6\pi)^-$		& \hfil --- \hfil  & $\geq 0.0052 $
\\
$\pi^-\pi^0\eta$ & $0.0101\pm0.0013$ & $0.0072\pm0.0011$ 
\\
$K^-K^0$ & $0.0089\pm0.0013$ & $0.0062\pm0.0016$   
\\
$\pi^-\phi$ & $< 0.002$ & $< 0.0006$
\\ \hline
\end{tabular}
\end{table}
%

The most recent results \cite{EI:96} are compared
with the $\tau$--decay measurements in Table~\ref{tab:vector}.
The agreement is quite good.
Moreover, the experimental precision of the $\tau$--decay data is
already better than the $e^+e^-$ one. 

The exclusive $\tau$ decays into final hadronic states with $J^P = 1^+$,
or Cabibbo suppressed  modes  with $J^P =1^-$,
cannot  be  predicted  with  the  same degree  of
confidence. We can only make model--dependent estimates \cite{PI:89}
with an accuracy which
depends on our ability to handle the strong interactions at low energies.
That just indicates that the decay of the $\tau$ lepton is
providing new experimental hadronic information.  
Owing to their semileptonic character, the hadronic $\tau$ decay data
are a unique and extremely useful tool to learn
about the couplings of the low--lying mesons to the weak currents.
 
\subsection{Chiral Dynamics}

At low momentum transfer, the coupling of any 
number of $\pi $'s, $K$'s and $\eta$'s to the
V$-$A current can be rigorously calculated with
Chiral Perturbation Theory techniques \cite{WE:79,GL:85,chpt:95,EC:95}.
In the absence of quark masses the QCD Lagrangian splits into two
independent chirality (left/right) sectors, with their own quark
flavour symmetries.
With three light quarks ($u$, $d$, $s$), the QCD Lagrangian
is then approximately
invariant under chiral $SU(3)_L\otimes SU(3)_R$ rotations
in flavour space. The vacuum is however not symmetric under the
chiral group. Thus, 
chiral symmetry breaks down to the usual eightfold--way $SU(3)_V$,
generating the appearance of eight Goldstone bosons in the
hadronic spectrum, which can be identified with the lightest
pseudoscalar octet; their small masses being generated by the 
quark mass matrix, which explicitly breaks chiral symmetry.
The Goldstone nature of the pseudoscalar octet implies strong
constraints on their low--energy interactions, which can be worked out
through an expansion in powers of momenta over the 
chiral symmetry--breaking scale \cite{WE:79,GL:85,chpt:95,EC:95}.

In the low--energy effective chiral realization of QCD, the vector
and axial--vector currents take the form \cite{PI:89,chpt:95}:
\beqn\label{eq:V_chpt}
V_\mu &\!\!\! = &\!\!\! -i \left(\Phi\lrder_\mu\Phi\right)
+ \cO(\Phi^4) + \cO(p^3)
- {i N_C\over 6\sqrt{2}\pi^2 f_\pi^3}\,\varepsilon_{\mu\nu\alpha\beta}
\, \left\{\partial^\nu\Phi\, \partial^\alpha\Phi\, \partial^\beta\Phi
+ \cO(\Phi^5)\right\} \, ,
\no\\ 
A_\mu &\!\!\! = &\!\!\! -\sqrt{2} f_\pi\,\partial_\mu\Phi
+ {\sqrt{2}\over 3 f_\pi}\,\left[\Phi,\left(\Phi\lrder_\mu\Phi\right)\right]
+ \cO(\Phi^5) + \cO(p^3)
\\ && \mbox{}
+ {N_C\over 12\pi^2 f_\pi^4}\,\varepsilon_{\mu\nu\alpha\beta}\,\left\{
\partial^\nu\Phi\,\partial^\alpha\Phi\,\left(\Phi
\stackrel{\leftrightarrow}{\partial^\beta}\Phi\right)
+ \cO(\Phi^6)\right\} \, , \no
\eeqn
where the odd-parity pieces, proportional to the Levi--Civita
pseudotensor, are generated by the Wess--Zumino--Witten term of the chiral
Lagrangian \cite{WZ:71,WI:83}, which incorporates the non-abelian chiral
anomaly of QCD.
The $3\times 3$ matrix
\bel{eq:Phi_matrix}
\Phi (x) \equiv {\vec{\lambda}\over\sqrt 2} \, \vec{\phi}
 \, = \, 
\pmatrix{{1\over\sqrt 2}\pi^0 \, + 
\, {1\over\sqrt 6}\eta_8
 & \pi^+ & K^+ \cr
\pi^- & - {1\over\sqrt 2}\pi^0 \, + \, {1\over\sqrt 6}\eta_8   
 & K^0 \cr K^- & \bar K^0 & - {2 \over\sqrt 6}\eta_8 }\, .
\ee
parametrizes the pseudoscalar octet fields. Thus,
at lowest order in momenta,
the couplings of the Goldstones to the weak current
can be calculated in a straightforward way.

The one--loop corrections are known \cite{GL:85,chpt:95,EC:95,CFU:96}
for the lowest--multiplicity
states ($\pi$, $K$, $2\pi$, $K\bar K$, $K\pi$, $3\pi$). 
Moreover, a two--loop calculation 
for the $2\pi$ decay mode is already available \cite{CFU:96}.
Therefore, exclusive hadronic $\tau$ decay data at low values of $q^2$
could be compared with rigorous QCD predictions.
There are also well--grounded theoretical results
(based on a $1/M_\rho$ expansion) for decays such as 
$\tau^-\to\nu_\tau(\rho\pi)^-,\nu_\tau(K^*\pi)^-,\nu_\tau(\omega\pi)^-$, 
but only in the
kinematical configuration where the pion is soft \cite{Wise}.

$\tau$ decays involve, however, high values of momentum transfer
where the chiral symmetry predictions no longer apply.
Since the relevant hadronic dynamics is governed by the non-perturbative
regime of QCD, we are unable at present to make first--principle
calculations for exclusive decays.
Nevertheless, one can still construct reasonable models, taking
into account the low--energy chiral theorems.
The simplest prescription \cite{PI:89,FWW:80,PI:87,GGP:90}
consists in extrapolating the chiral predictions
to higher values of $q^2$, by suitable final--state--interaction 
enhancements  which take into account the resonance structures
present in each channel in a phenomenological way.
This can be done weighting the contribution of a given set of
pseudoscalars, with definite quantum numbers, with an appropriate
resonance form factor such as
\bel{eq:F_model}
F_R(s) \, = \, {M_R^2\over M_R^2-s-i M_R\Gamma_R(s)} \, ,
\ee
where $M_R$ ($\Gamma_R$) denote the mass (width) of the resonance $R$.
The requirement that the chiral predictions must be recovered
below the resonance region fixes the normalization of those form factors
to be one at zero invariant mass.

The extrapolation of the low--energy chiral theorems provides
a useful description of the $\tau$ data in terms of a few resonance
parameters. 
Therefore, it has been extensively used 
\cite{KS:90,PI:89,FWW:80,PI:87,GGP:90,karlsruhe,DE:94}
to analyze the
main $\tau$ decay modes, and has been incorporated into
the TAUOLA Monte Carlo library \cite{tauola}.
However, the model is too naive to be considered as
an actual implementation of the QCD dynamics.
Quite often, the numerical predictions could be drastically changed
by varying some free parameter or modifying the form--factor
ansatz.
Not surprisingly, some predictions fail badly to reproduce the
experimental data whenever a new resonance structure shows up
\cite{Shelkov}.

The addition of resonance form factors to the chiral low--energy
amplitudes does not guarantee that the
chiral symmetry constraints on the
resonance couplings have been correctly implemented. 
The proper way of including higher--mass states into the effective 
chiral theory was developed in Refs.~\citenum{EGLPR:89}.
Using these techniques, a refined calculation of the rare decay
$\tau^-\to\nu_\tau\eta\pi^-$ has been given recently \cite{NR:95}.
A systematic analysis of $\tau$ decay amplitudes within this
framework is in progress \cite{GP:97}.

Tau decays offer a very good laboratory to improve our present
understanding of the low--energy QCD dynamics. 
The general form factors characterizing the non-perturbative
hadronic decay amplitudes
can be experimentally extracted from the
Dalitz--plot distributions of the final hadrons \cite{KM:92}.
An exhaustive analysis of $\tau$ decay modes
would provide a very valuable data basis to confront with
theoretical models.

\setcounter{equation}{0}
\section{QCD ANALYSIS OF THE TAU HADRONIC WIDTH}
\label{sec:inclusive_width}

The inclusive character of the total $\tau$ hadronic width renders possible 
an accurate calculation of the ratio
\cite{BR:88,NP:88,ORSAY:90,BNP:92,LDP:92a,OHIO:92,QCD:94,NA:95,BR:96,HOP:96}
[$(\gamma)$ represents additional photons or lepton pairs]
\be\label{eq:r_tau_def}
     R_\tau \equiv { \Gamma [\tau^- \rightarrow \nu_\tau
                   \,\mbox{\rm hadrons}\, (\gamma)] \over
                         \Gamma [\tau^- \rightarrow
                \nu_\tau e^- {\bar \nu}_e (\gamma)] } ,
\ee
using standard field theoretic methods. 

The theoretical analysis of $R_\tau$ involves
the two--point correlation functions
\beqn\label{eq:pi_v}
\Pi^{\mu \nu}_{ij,V}(q) &\!\!\!\equiv &\!\!\!
 i \int d^4x \, e^{iqx} 
\langle 0|T(V^{\mu}_{ij}(x) V^{\nu}_{ij}(0)^\dagger)|0\rangle \, ,
\\ 
\label{eq:pi_a}
\Pi^{\mu \nu}_{ij,A}(q) &\!\!\!\equiv &\!\!\!
 i \int d^4x \, e^{iqx} 
\langle 0|T(A^{\mu}_{ij}(x) A^{\nu}_{ij}(0)^\dagger)|0\rangle \, ,
\eeqn
for the vector $\, V^{\mu}_{ij} = \bar{\psi}_j \gamma^{\mu} \psi_i \, $
and axial--vector 
$\, A^{\mu}_{ij} = \bar{\psi}_j \gamma^{\mu} \gamma_5 \psi_i \,$
colour--singlet quark currents ($i,j=u,d,s$).
They have the Lorentz decompositions
\bel{eq:lorentz}
\Pi^{\mu \nu}_{ij,V/A}(q) \, = \,
 (-g^{\mu\nu} q^2 + q^{\mu} q^{\nu}) \, \Pi_{ij,V/A}^{(1)}(q^2)
 + q^{\mu} q^{\nu} \, \Pi_{ij,V/A}^{(0)}(q^2)\, ,
\ee
where the superscript $(J=0,1)$ 
denotes the angular momentum in the hadronic rest frame.
  
The imaginary parts of the two--point functions
$\, \Pi^{(J)}_{ij,V/A}(q^2) \, $ 
are proportional to the spectral functions for hadrons with the 
corresponding quantum numbers.  The hadronic decay rate of the $\tau$
can be written as an integral of these spectral functions
over the invariant mass $s$ of the final--state hadrons:
\bel{eq:spectral}
R_\tau   =  
12 \pi \int^{m_\tau^2}_0 {ds \over m_\tau^2 } \,
 \left(1-{s \over m_\tau^2}\right)^2 
\left[ \left(1 + 2 {s \over m_\tau^2}\right) 
 \mbox{\rm Im} \Pi^{(1)}(s)
 + \mbox{\rm Im} \Pi^{(0)}(s) \right] \, . 
\ee
The appropriate combinations of correlators are 
\bel{eq:pi}
\Pi^{(J)}(s)  \equiv
|V_{ud}|^2 \, \left[ \Pi^{(J)}_{ud,V}(s) + \Pi^{(J)}_{ud,A}(s) \right]
+ |V_{us}|^2 \, \left[ \Pi^{(J)}_{us,V}(s) + \Pi^{(J)}_{us,A}(s) \right]
\, . 
\ee
%

\begin{figure}[thb]
\vfill
\centerline{
\begin{minipage}[t]{.47\linewidth}\centering
\centerline{\epsfxsize =7.5cm \epsfbox{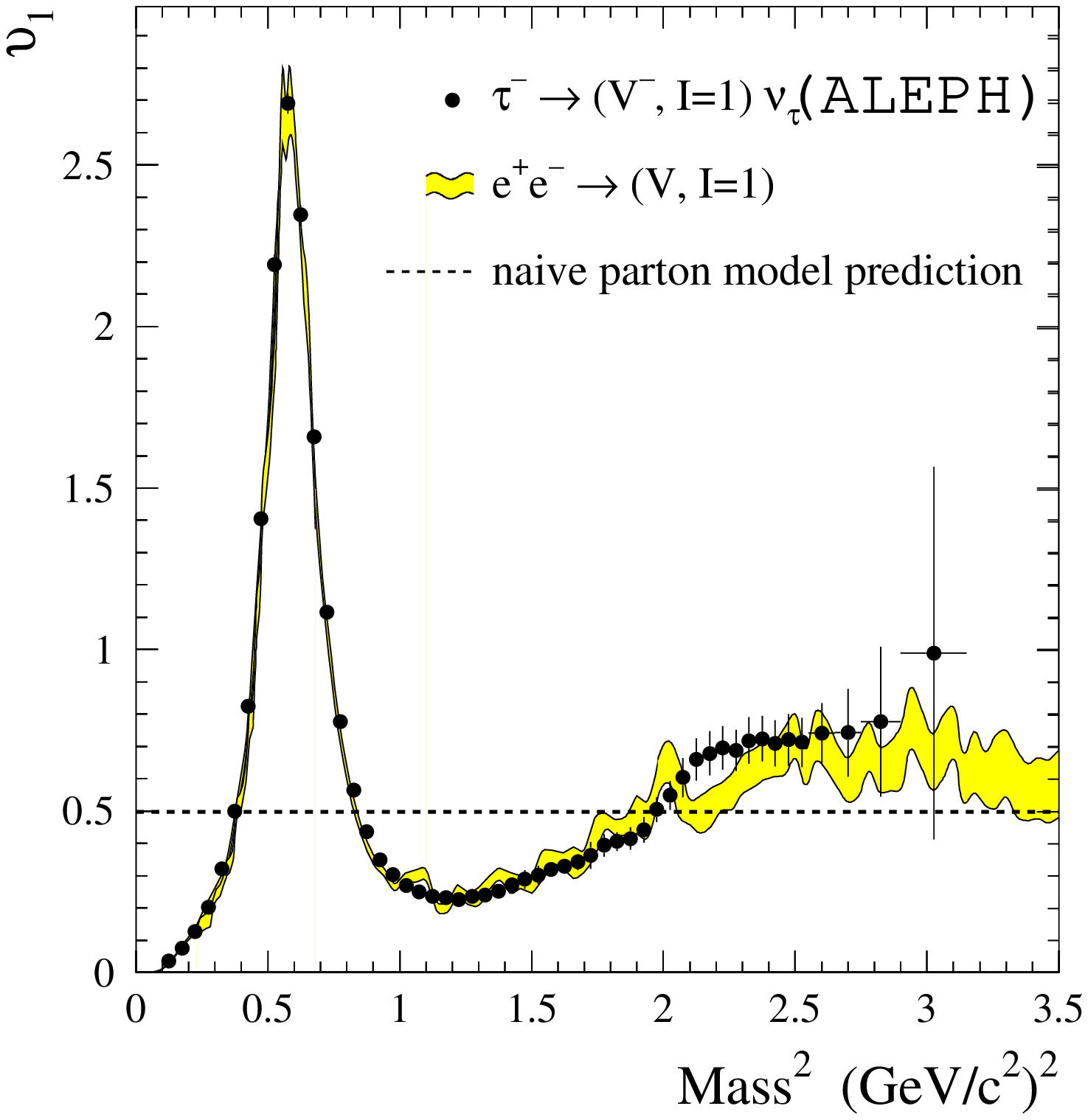}}
\caption{Measured vector--current spectral function \protect\cite{Hocker}
[$v_1(s) \equiv 2\pi$ Im$\Pi^{(1)}_{ud,V}(s)$].
The points correspond to $\tau$ decay data whereas the band has been
obtained from $e^+e^-$. The dashed line represents the naive
parton model prediction.
(Taken from Ref.~\protect\citenum{Hocker}).
}
\label{fig:vsf}
\end{minipage}
\hspace{0.67cm}
\begin{minipage}[t]{.47\linewidth}\centering
\centerline{\epsfxsize =7.7cm \epsfbox{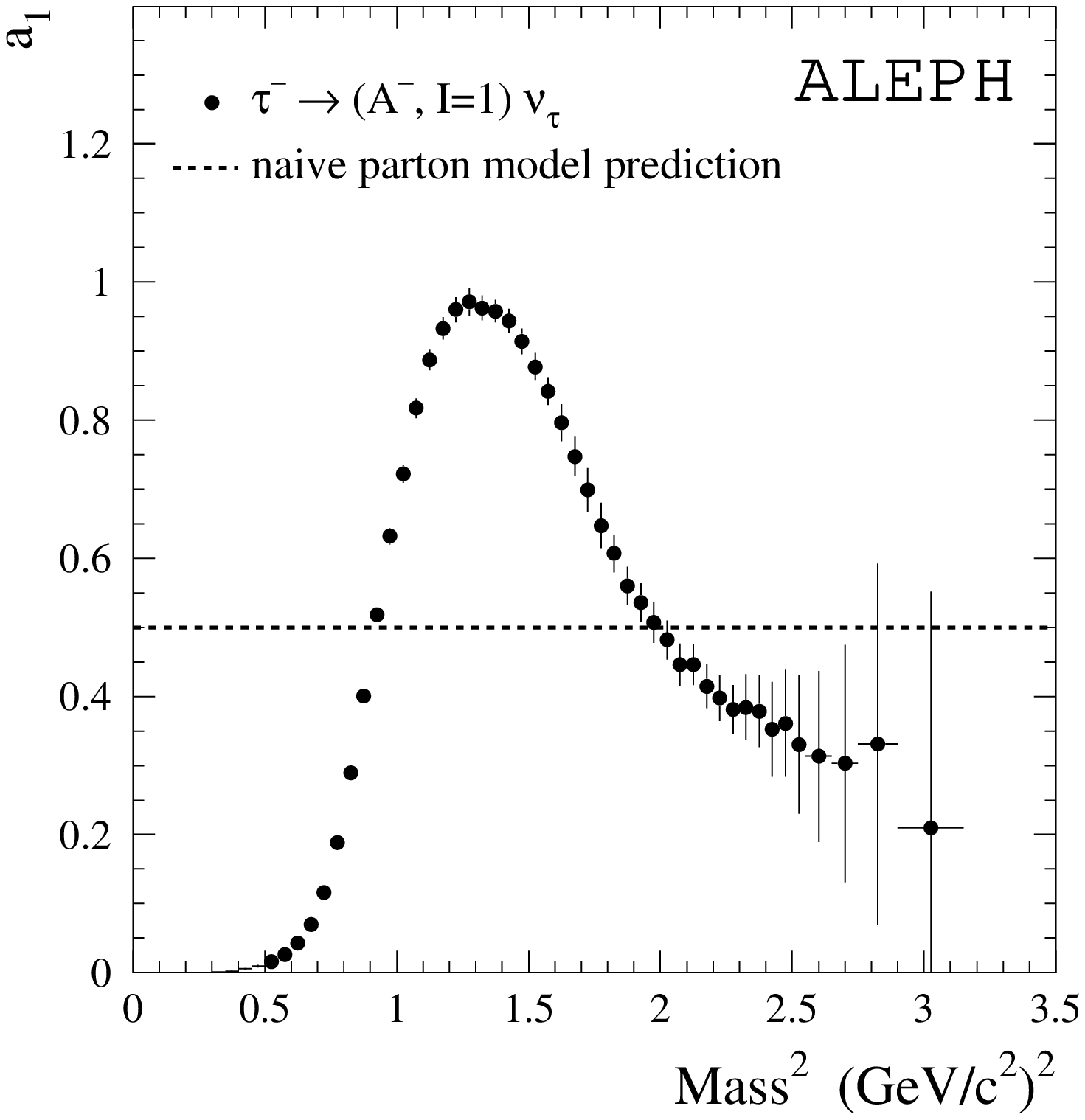}}
\caption{Measured axial--current spectral function \protect\cite{Hocker}
[$a_1(s) \equiv 2\pi$ Im$\Pi^{(1)}_{ud,A}(s)$].
The dashed line represents the naive parton model prediction.
(Taken from Ref.~\protect\citenum{Hocker}).
}
\label{fig:asf}
\end{minipage}
}
\vfill
\end{figure}

We can separate the inclusive contributions associated with
specific quark currents:
\be\label{eq:r_tau_v,a,s}
 R_\tau \, = \, R_{\tau,V} + R_{\tau,A} + R_{\tau,S}\, .
\ee
$R_{\tau,V}$ and $R_{\tau,A}$ correspond to the first two terms
in \eqn{eq:pi}, while  
$R_{\tau,S}$ contains the remaining Cabibbo--suppressed contributions.
Non-strange hadronic decays of the $\tau$ are resolved experimentally
into vector ($R_{\tau,V}$) and axial--vector ($R_{\tau,A}$)
contributions according to whether the
hadronic final state includes an even or odd number of pions.
Strange decays ($R_{\tau,S}$) are of course identified by the
presence of an odd number of kaons in the final state.

Since the hadronic spectral functions are sensitive to the non-perturbative
effects of QCD that bind quarks into hadrons, the integrand in 
Eq.~\eqn{eq:spectral} cannot be calculated at present from QCD.
Nevertheless the integral itself can be calculated systematically
by exploiting
the analytic properties of the correlators $\Pi^{(J)}(s)$.
They are analytic
functions of $s$ except along the positive real $s$ axis, where their
imaginary parts have discontinuities.  The integral \eqn{eq:spectral} can
therefore be expressed as a contour integral 
in the complex $s$ plane running
counter--clockwise around the circle $|s|=m_\tau^2$:
\bel{eq:circle}
 R_\tau  = 
6 \pi i \oint_{|s|=m_\tau^2} {ds \over m_\tau^2} \,
 \left(1 - {s \over m_\tau^2}\right)^2
 \left[ \left(1 + 2 {s \over m_\tau^2}\right) \Pi^{(0+1)}(s)
         - 2 {s \over m_\tau^2} \Pi^{(0)}(s) \right] \, . 
\ee
%

\begin{figure}[htb]
\centerline{\epsfxsize =8cm \epsfbox{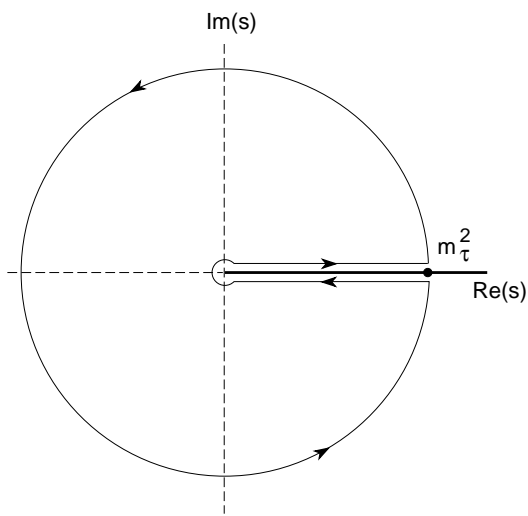}}
\label{fig:circle}
\caption{Integration contour in the complex s plane, used to obtain
Eq.~\protect\eqn{eq:circle}.}
\end{figure}

The advantage of expression \eqn{eq:circle}  
over \eqn{eq:spectral} for $R_\tau$
is that it requires the correlators only for
complex $s$ of order $m_\tau^2$, which is significantly larger than 
the scale
associated with non-perturbative effects in QCD.  The short--distance
Operator Product Expansion (OPE) can therefore be used to organize
the perturbative and non-perturbative contributions
to the correlators into a systematic expansion \cite{SVZ:79}
in powers of $1/s$,
\be\label{eq:ope}
 \Pi^{(J)}(s) = \sum_{D=2n}\,\sum_{\mbox{\rms dim} {\cal O} = D}
 {{\cal C}^{(J)}(s,\mu) \,\langle {\cal O}(\mu)\rangle\over (-s)^{D/2}} ,
\ee
 where the inner sum is over local gauge--invariant
 scalar operators of dimension $D=0,2,4,\ldots $
 The possible uncertainties associated with the use of the OPE near the
 time--like axis are negligible in this case, because
 the integrand in Eq.~\eqn{eq:circle} includes a factor
 $(1- s/m_\tau^2)^2$, which provides a double zero at $s=m_\tau^2$,
 effectively suppressing the contribution from the
 region near the branch cut.
 The parameter $\mu$ 
 is an arbitrary factorization scale, which separates long--distance
 non-perturbative effects, which are absorbed into the vacuum matrix elements
 $\langle {\cal O}(\mu)\rangle $, from short--distance effects, which belong
 in the Wilson coefficients ${\cal C}^{(J)}(s,\mu)$.
 The $D=0$ term (unit operator) corresponds to the pure perturbative
contributions, neglecting quark masses. The leading quark--mass
corrections     generate the $D=2$ term. The first dynamical operators
involving non-perturbative physics appear at $D=4$.
Inserting the functions \eqn{eq:ope}
into \eqn{eq:circle} and evaluating the contour integral, $R_\tau$
can be expressed as an expansion in powers of $1/m_\tau^2$,
with coefficients that depend only logarithmically on $m_\tau$.

It is convenient to express the corrections to $R_\tau$
from dimension--$D$ operators in terms of the
fractional corrections $\delta^{(D)}_{ij,V/A}$ to the 
naive contribution
from the current with quantum numbers $ij,V$ or $ij,A$:
\be\label{eq:r_v}
R_{\tau,V/A}  =  {3 \over 2} |V_{ud}|^2 
   S_{EW} \left( 1 + \delta_{EW}' + \delta^{(0)} + 
      \sum_{D=2,4,...} \delta^{(D)}_{ud,V/A} \right)  , 
\ee
\be \label{eq:r_s} 
R_{\tau,S}  = 
 3 |V_{us}|^2 S_{EW} \left( 1 + \delta_{EW}' + \delta^{(0)} +
  \sum_{D=2,4,...} \delta^{(D)}_{us} \right) ,
\ee
where 
$\delta^{(D)}_{ij} = (\delta^{(D)}_{ij,V} + \delta^{(D)}_{ij,A})/2$
is the average of the vector and axial--vector corrections,
and the factors \cite{MS:88}
\bel{eq:s_ew}
 S_{EW}  = 
\left( {\alpha(m_b^2) \over \alpha(m_\tau^2)} \right)^{9\over19}
         \left( {\alpha(M_W^2) \over \alpha(m_b^2)} \right)^{9\over 20}
         \left( {\alpha(M_Z^2) \over \alpha(M_W^2)} \right)^{36\over 17}
      =  1.0194  
\ee
and  \cite{BL:90}
\bel{eq:d_ew'}
 \delta_{EW}' = {5\over 12} \, {\alpha(m_\tau^2)\over\pi } \simeq 0.0010 \, ,
\ee
contain the known electroweak corrections. 

Adding the three terms, the total ratio $R_\tau$ is
\bel{eq:r_total}
R_{\tau} = 
 3 \left( |V_{ud}|^2 + |V_{us}|^2 \right)
S_{EW} \biggl\{ 1 + \delta_{EW}' + \delta^{(0)} 
+ \sum_{D=2,4,...} \left( \cos^2 \theta_C \delta^{(D)}_{ud}
 + \sin^2 \theta_C \delta^{(D)}_{us} \right) \biggr\} ,                                         
\ee
where 
$\sin^2\theta_C\equiv |V_{us}|^2/(|V_{ud}|^2 + |V_{us}|^2)$.

\subsection{Perturbative corrections}

The dimension--0 contribution,
$\delta^{(0)} = \delta^{(0)}_{ij,V/A}$,
is the purely perturbative correction neglecting quark masses,
which is the same for all the components of $R_\tau$.
It is given by
\cite{BR:88,NP:88,ORSAY:90,BNP:92,LDP:92a,OHIO:92,QCD:94}:
\be\label{eq:delta0}
\delta^{(0)}  = \sum_{n=1}  K_n \, A^{(n)}(\alpha_s) 
=  
{\alpha_s(m_\tau^2)\over\pi}
  + 5.2023 \left( {\alpha_s(m_\tau^2)\over\pi}\right)^2
 + 26.366 \left( {\alpha_s(m_\tau^2)\over\pi}\right)^3
       + \, \cO(\alpha_s^4)  \, .
\ee

The dynamical coefficients $K_n$ regulate the perturbative expansion
of the correlator $D(s)\equiv -s{d\over ds}\Pi^{(0+1)}(s)$ 
in the massless--quark limit
[$s\Pi^{(0)}(s)=0$ for massless quarks]; they are
known \cite{ChKT:79,GKL:91,SS:91} to $\cO(\alpha_s^3)$:
$K_1 = 1$; $K_2 = 1.63982$; $K_3(\overline{MS}) = 6.37101$.
The kinematical effect of the contour integration is contained in
the functions \cite{LDP:92a}
\bel{eq:a_xi}
A^{(n)}(\alpha_s)  =  {1\over 2 \pi i}
\oint_{|s| = m_\tau^2} {ds \over s} \,
  \left({\alpha_s(-s)\over\pi}\right)^n  
 \left( 1 - 2 {s \over m_\tau^2} + 2 {s^3 \over m_\tau^6}
         - {s^4 \over  m_\tau^8} \right) 
= \left({\alpha_s(m_\tau^2)\over\pi}\right)^n + \cO(\alpha_s^{n+1})  \, ,
\ee
which only depend on $\alpha_s(m_\tau^2)$.
Owing to the long running of the strong coupling along the circle, the
coefficients of the
perturbative expansion of $\delta^{(0)}$ in powers of
$\alpha_s(m_\tau^2)$ are larger than the direct $K_n$
contributions. This running effect can be properly resummed to all orders
in $\alpha_s$ by fully keeping \cite{LDP:92a}
the known three--loop--level calculation of
the integrals $A^{(n)}(\alpha_s)$. 

%
\begin{table}[thb]
\centering
\caption{$\delta^{(0)}$ for different values of $\alpha_s(m_\tau^2)$.}
\label{tab:perturbative}
\vspace{0.4cm}
\begin{tabular}{|c|c|c|c|}            
\hline  
$\alpha_s(m_\tau^2)$ & \multicolumn{2}{c|}{$\delta^{(0)}$} &
$\Delta(\delta^{(0)})$ 
\\ \cline{2-3}
& $K_4 = 0$ & $K_4=27.5$ &
\\ \hline
$0.30$ & $0.161$ & $0.164$ & $\pm0.006$
\\
$0.31$ & $0.168$ & $0.172$ & $\pm0.007$
\\
$0.32$ & $0.176$ & $0.180$ & $\pm0.008$
\\
$0.33$ & $0.183$ & $0.188$ & $\pm0.008$
\\
$0.34$ & $0.191$ & $0.196$ & $\pm0.009$
\\
$0.35$ & $0.198$ & $0.203$ & $\pm0.010$
\\
$0.36$ & $0.205$ & $0.211$ & $\pm0.010$
\\       
$0.37$ & $0.213$ & $0.219$ & $\pm0.011$
\\
$0.38$ & $0.220$ & $0.226$ & $\pm0.012$
\\
$0.39$ & $0.227$ & $0.234$ & $\pm0.012$
\\
$0.40$ & $0.234$ & $0.241$ & $\pm0.013$
\\  \hline   
\end{tabular}
\end{table}
%

The dominant perturbative uncertainties come  from
the unknown higher--order coefficients $K_{n>3}$.
The $\cO(\alpha_s^4)$ contribution has been estimated \cite{KS:95}
using scheme--invariant methods, namely 
the principle of minimal sensitivity \cite{ST:81}
and the effective charge approach \cite{GR:80}, with the
result \cite{KS:95}:
\bel{eq:k4}
K_4^{\mbox{\rms est}} = 27.5 \, .
\ee
This number is very close to the na\"{\i}ve guess \cite{LDP:92a}
$K_4 \sim (K_3/K_2) K_3 \approx 25$.
A similar estimate,
$K_4^{\mbox{\rms NNA}} = 24.8$, 
is obtained 
\cite{BR:93,BRK:93,BE:93,LTM:94,BB:95,BBB:95,NE:96}
in the limit of a large number of quark
flavours, using the so-called {\it naive non-abelianization}
prescription \cite{BG:95}
($n_f \to 3\beta_1 = n_f -{33\over 2} = -{27\over 2}$).
From a fit to the experimental $\tau$ data, the value
$K_4^{\mbox{\rms fit}}=29\pm 5$ has been also quoted \cite{LD:94}.

Using the estimate \eqn{eq:k4}, the $\cO(\alpha_s^4)$ correction
amounts to a 0.004 increase of $\delta^{(0)}$ for $\alpha_s(m_\tau^2)=0.35$.
The resulting perturbative contribution $\delta^{(0)}$
is given \cite{HOP:96} in
Table~\ref{tab:perturbative} for different values of 
the strong coupling constant $\alpha_s(m_\tau^2)$.
In order to be conservative, and to account for all possible sources
of perturbative uncertainties,
we have  used \cite{OHIO:92,QCD:94,HOP:96}
\bel{eq:perror} 
\Delta(\delta^{(0)}) = \pm 50 \, A^{(4)}(\alpha_s)\, ,
\ee
as an estimate of the theoretical error on $\delta^{(0)}$.
Note that, for the relevant values of $\alpha_s$, this is
of the same size as $K_3\, A^{(3)}(\alpha_s)$; thus, this
error estimate is conservative enough to apply \cite{QCD:94}
in the worst possible scenario,
where the onset of the asymptotic behaviour of the
perturbative series were already reached for $n=3,4$.

There have been attempts
\cite{BE:93,LTM:94,BB:95,BBB:95,NE:96}
to improve the perturbative prediction
by performing an all--order summation a certain class
of higher--order corrections (the so-called ultraviolet
renormalon chains). This can be accomplished using
exact large--$n_f$ results and applying the
{\it naive non-abelianization} prescription \cite{BG:95}.
Unfortunately, the naive resummation turns out to be renormalization--scheme
dependent beyond one loop \cite{LTM:94,CH:95}.
More recently, a renormalization--scheme--invariant summation has been
presented \cite{MT:96}.
The final effect of the higher--order corrections (beyond $K_4$)
turns out to be small.

\subsection{Power corrections}

The leading quark--mass corrections $\delta^{(2)}_{ij}$ are known
\cite{BNP:92,CK:93} to order $\alpha_s^2$.
They are certainly tiny for the up and down quarks
($\delta^{(2)}_{ud}\sim -0.08\% $), but 
the correction from the strange quark mass is important
for strange decays
($\delta^{(2)}_{us}\approx  -20\% $).
Nevertheless, because of the $\sin^2{\theta_C}$ suppression, the
effect on the total ratio $R_{\tau}$  is only $-(1.0\pm 0.2) \%$.

The leading non-perturbative contributions can be shown to be
suppressed by six powers of the $\tau$ mass 
\cite{BR:88,NP:88,ORSAY:90,BNP:92},
and are therefore very small.
This fortunate fact is due to the phase--space factors in
\eqn{eq:circle}; their form is such that the leading $1/s^2$ corrections
to $\Pi^{(1)}(s)$ do not survive the integration along the circle.
Moreover, there is a large cancellation 
between the vector and axial--vector $D=6$ contributions
to the total hadronic width
(the $D=6$ operator with the largest Wilson coefficient contributes with
opposite signs to the vector and axial--vector correlators, due to the
$\gamma_5$ flip). Thus, the
non-perturbative corrections to $R_\tau$ are smaller than the
corresponding contributions to  $R_{\tau,V/A}$.

The numerical size of the non-perturbative corrections can be 
determined from the invariant--mass distribution of the final hadrons
in $\tau$ decay \cite{PI:89}.
Although the distributions themselves cannot be predicted at present,
certain weighted integrals of the hadronic spectral functions can be
calculated in the same way as $R_\tau$.
The analyticity properties of $\Pi^{(J)}_{ij,V/A}$ imply 
\cite{PI:89,LDP:92b}:
%
\bel{eq:weighted_integrals}
\int_0^{s_0} ds\, W(s)\: \mbox{\rm Im}\Pi^{(J)}_{ij,V/A}\, =\,
{i\over 2} \oint_{|s|=s_0} ds\, W(s) \,\Pi^{(J)}_{ij,V/A}\, , 
\ee
with $W(s)$ an arbitrary weight function without singularities in the
region $|s|\leq s_0$.
Generally speaking, the accuracy of the theoretical predictions can be
much worse than the one of $R_\tau$, because non-perturbative effects
are not necessarily suppressed.
In fact, choosing an appropriate weight function, non-perturbative effects
can even be made to dominate the final result. But this is precisely
what makes these integrals interesting: they can be used to measure the
parameters characterizing the non-perturbative dynamics.

To perform an experimental analysis, it is convenient to use
moments of the directly measured invariant--mass distribution
\cite{LDP:92b} ($k,l\ge 0$)
\be\label{eq:moments}
R^{kl}_\tau(s_0) \equiv\int_0^{s_0}\, ds
 \, \left(1 - {s\over s_0}\right)^k \left ( {s \over m_\tau^2} \right )^l
{ d R_\tau \over d s} \, .
\ee
The factor $(1-s/s_0)^k$ supplements $(1-s/m_\tau^2)^2$ for 
$s_0\not= m_\tau^2$,
in order to squeeze the integrand at the crossing of the positive real axis
and, therefore, improves the reliability of the OPE analysis; moreover, for
$s_0=m_\tau^2$ it reduces the contribution from the tail of the distribution,
which is badly defined experimentally.
A combined fit of
different $R_\tau^{kl}(s_0)$ moments
results in experimental
values for $\alpha_s(m_\tau^2)$ 
and for the coefficients of the inverse power corrections
in the OPE.
$R_\tau^{00}(m_\tau^2) = R_\tau$ 
uses the overall normalization of the hadronic
distribution, while the ratios 
$D_\tau^{kl}(m_\tau^2) = R^{kl}_\tau (m_\tau^2)/R_\tau$ are based on
the shape of the $s$ distribution and are more dependent on 
non-perturbative effects \cite{LDP:92b}.

The predicted suppression \cite{BR:88,NP:88,ORSAY:90,BNP:92}
of the non-perturbative corrections has been confirmed by
ALEPH \cite{Hocker,ALEPH:93,DU:95} and CLEO \cite{CLEO:95},
using the moments (0,0), (1,0), (1,1), (1,2) and (1,3).
The most recent ALEPH analysis \cite{ALEPH:97b} gives:
\bel{eq:del_np}
\delta_{\mbox{\rms NP}} \equiv \sum_{D\geq 4} 
\left( \cos^2 \theta_C \delta^{(D)}_{ud}
         + \sin^2 \theta_C \delta^{(D)}_{us} \right)
=     -(0.02\pm 0.5)\% \, ,      
\ee
in agreement with previous estimates \cite{BNP:92}.

\goodbreak
\subsection{Phenomenology}

The QCD prediction
for $R_\tau$ is then completely dominated by the
perturbative contribution $\delta^{(0)}$; 
non-perturbative effects being of the order of
the perturbative uncertainties from uncalculated higher--order
corrections \cite{QCD:94,NA:95,BR:96,HOP:96}. Furthermore, 
as shown in Table~\ref{tab:perturbative}, the result turns out to be
very sensitive to the value of $\alpha_s(m_\tau^2)$, allowing for an 
accurate determination of the fundamental QCD coupling.

The experimental value for $R_\tau$ can be obtained
from the leptonic branching fractions or from the
$\tau$ lifetime. The average of those determinations \cite{PI:96}
\be  
  R_\tau =3.649 \pm 0.014 \, ,
\ee
corresponds to 
\be\label{eq:alpha}
\alpha_s(m_\tau^2)  =  0.35\pm 0.02 \, . 
\ee

Once the running coupling constant $\alpha_s(s)$ is determined at the scale
$m_\tau$, it can be evolved to higher energies using the renormalization
group.  The size of its error bar scales roughly
as $\alpha_s^2$, and it therefore shrinks as the scale increases.
Thus a modest precision in the
determination of $\alpha_s$ at low energies results in a very high
precision in the coupling constant at high energies.
After evolution up to the scale $M_Z$, the strong coupling constant in
\eqn{eq:alpha} decreases to
%
\be\label{eq:alpha_z}
\alpha_s(M_Z^2)  =  0.1217\pm 0.0025 \, ,
\ee
in excellent
agreement with the direct measurement \cite{SC:97,BE:96} at $\mu=M_Z$,
$\alpha_s(M_Z^2)  =  0.121\pm0.003$,  
and with a similar error bar.
The comparison of these two determinations of $\alpha_s$ in two extreme
energy regimes, $m_\tau$ and $M_Z$, provides a beautiful test of the
predicted running of the QCD coupling.

\begin{figure}[tbh]
\centerline{\epsfxsize =7cm \epsfbox{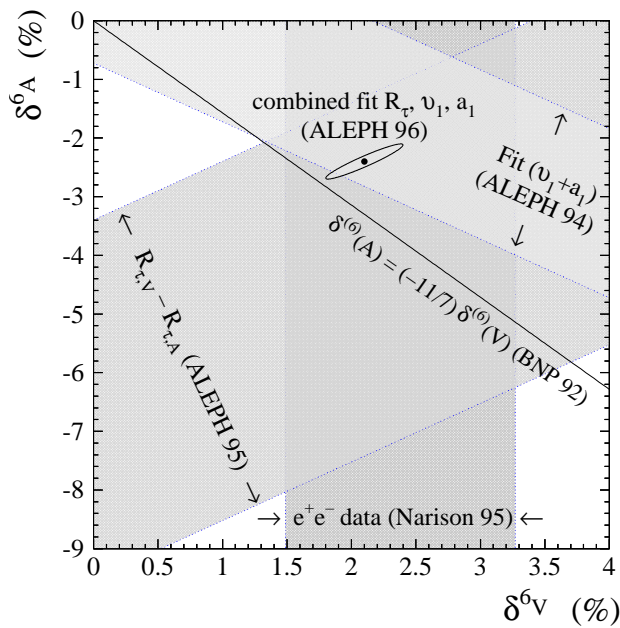}}
\vspace{-1.0cm}
\caption{Constraints on 
$\delta^{(6)}_V$ and $\delta^{(6)}_A$ obtained from
ALEPH data \protect\cite{Hocker}.
The ellipse depicts the combined fit.
All results are still preliminary. 
(Taken from Ref.~\protect\citenum{Hocker}).\hfill\mbox{}}
\label{fig:delta_6}
\end{figure}

With $\alpha_s(m_\tau^2)$ fixed to the value in Eq.~(\ref{eq:alpha}), 
the same theoretical framework gives definite
predictions \cite{BNP:92,QCD:94} for the semi-inclusive $\tau$ decay widths
$R_{\tau,V}$, $R_{\tau,A}$ and $R_{\tau,S}$, in good agreement with the
experimental measurements \cite{Hocker,Davier}.
The separate analysis of the vector and axial--vector contributions
allows to investigate the associated non-perturbative corrections.
Figure~\ref{fig:delta_6} shows \cite{Hocker}
the (preliminary) constraints on 
$\delta^{(6)}_V$ and $\delta^{(6)}_A$ obtained from the most recent
ALEPH analyses \cite{ALEPH:93,Hocker,ALEPH:96}.
A clear improvement over previous phenomenological determinations 
\cite{BNP:92,NA:95b} is apparent.

The Cabibbo--suppressed width $R_{\tau,S}$ is very
sensitive to the value of the strange quark mass \cite{BNP:92},
providing a direct and clean way of measuring $m_s$.
A very preliminary value,
$m_s(m_\tau^2)= (212\, {}^{+30\, +1}_{-35\, -5})$ MeV,
has been already presented at the
last $\tau$ workshop \cite{Davier}.

Using the measured invariant--mass distribution of the final hadrons, it
is possible to evaluate the integral $R_\tau^{00}(s_0)$,
with an arbitrary
upper limit of integration $s_0\leq m_\tau^2$. The experimental $s_0$
dependence  agrees well with the
theoretical predictions \cite{LDP:92b} up to rather low values\cite{GN:96}
of $s_0$ ($> 0.7$ GeV$^2$). 
Equivalently,  from the measured \cite{DU:95,CLEO:95}
$R_\tau^{00}(s_0)$ distribution one obtains
$\alpha_s(s_0)$ as a function of the scale $s_0$.
As shown  \cite{GN:96} in Figure~\ref{fig:alphang},
the result exhibits an impressive agreement with the
running predicted at three--loop order by QCD.
It is important to realize \cite{GN:96}
that the theoretical prediction for
$R_\tau^{00}(s_0)$ does not contain inverse powers of $s_0$
(as long as the $s$--dependence of the Wilson coefficients is ignored).
The power corrections are suppressed by powers of $1/m_\tau^2$;
thus, they do not drive a break--down of the OPE. This could explain
the surprisingly good agreement with the data for $s_0\lsim 1$ GeV$^2$.

\begin{figure}[bht]
\centerline{\epsfxsize =9.5cm \epsfbox{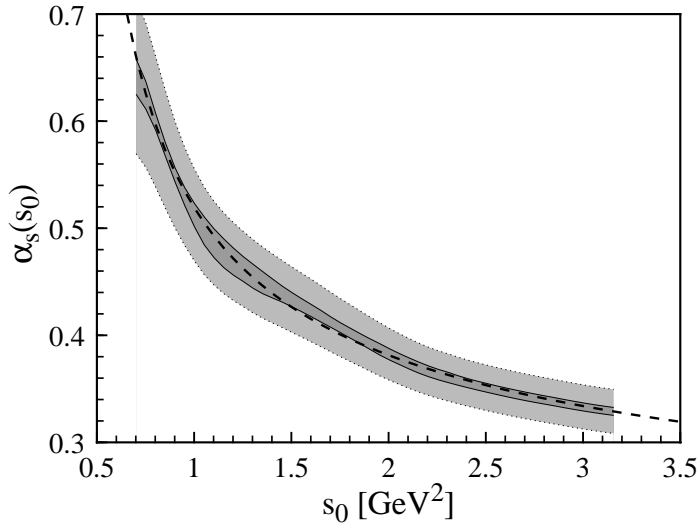}}
\caption{Values of $\alpha_s(s_0)$ extracted 
\protect\cite{GN:96} from the $R_\tau^{00}(s_0)$ data 
\protect\cite{DU:95,CLEO:95}. 
The dashed line shows the three--loop QCD
prediction for the running coupling constant.
(Taken from Ref.~\protect\citenum{GN:96}).\hfill\mbox{}}
\label{fig:alphang}
\end{figure}

A similar test was performed before \cite{NP:93} for $R_{\tau,V}$, 
using the vector spectral
function measured in $e^+e^-\to$ hadrons, and
varying the value of the $\tau$ mass. This allows to
study the behaviour of the OPE at lower scales.
The theoretical predictions for $R_{\tau,V}$ as function of $m_\tau^2$
agree \cite{NP:93} well with the data for $m_\tau > 1.2$ GeV.
Below this value, higher--order inverse power corrections become
very important and eventually generate the expected break--down
of the expansion in powers of $1/m_\tau^2$.


  The theoretical analysis of the $\tau$ hadronic width has reached a very
mature level.
Many different sources of possible perturbative and non-perturbative
contributions have been analyzed.
A very detailed study of the associated
uncertainties has been given in Ref.~\citenum{QCD:94}.
%
The comparison of the theoretical  predictions with the experimental
data shows a successful and consistent picture.
The resulting $\alpha_s(m_\tau^2)$ determination is in excellent agreement with
(and more precise than)
the measurements at the $M_Z$ scale, 
providing clear evidence of the running of
$\alpha_s$. The QCD predictions are 
further confirmed by
analyses of the semi-inclusive components
of the  $\tau$ hadronic width, $R_{\tau,V}$, $R_{\tau,A}$ 
and $R_{\tau,S}$, and the invariant-mass distribution of the final
decay products.

 In addition to provide beautiful tests of perturbative QCD, 
the hadronic spectral
functions measured in $\tau$ decay contain valuable dynamical information
on non-perturbative aspects of the strong interactions \cite{PI:89}
which could greatly enhance our present understanding of these phenomena.
For instance, $R_{\tau,V}-R_{\tau,A}$ is a pure non-perturbative
quantity; basic QCD properties force
the associated invariant--mass distribution to obey a
series of chiral sum rules 
\cite{PI:89,WE:67,DGMLY:67,FNR:79,PS:87,KaMa:92,DG:94},
which have been recently tested with $\tau$ data \cite{Hocker}.
The measurement of the vector spectral function \cite{ALEPH:97}
Im$\Pi_V(s)$ has also been used \cite{ADH:97}
 to reduce the present uncertainties in fundamental
QED quantities such as $\alpha(M_Z)$ and $(g^\gamma_\mu-2)$.

\section{SUMMARY}
\label{sec:summary}

The flavour structure of the Standard Model
 is one of the main pending questions
in our understanding of weak interactions. Although we do not know the
reason of the observed family replication, we have learned experimentally
that the number of Standard Model fermion generations is just three 
(and no more).
Therefore, we must study as precisely as possible the few existing flavours
to get some hints on the dynamics responsible for their observed structure.

The $\tau$ turns out to be an ideal laboratory to test the Standard Model. 
It is a lepton, which means clean physics, and moreover it is
heavy enough to produce a large variety of decay modes.
Na\"{\i}vely, one would expect the $\tau$ to be much more sensitive
than the $e$ or the $\mu$ to new physics related to the flavour and
mass--generation problems.

QCD studies can also benefit a lot from the existence of this heavy lepton,
able to decay into hadrons. Owing to their semileptonic character, the
hadronic $\tau$ decays provide a powerful tool to investigate the low--energy
effects of the strong interactions in rather simple conditions.

Our knowledge of the $\tau$ properties has been considerably
improved during the last few years. 
Lepton universality has been tested to rather good accuracy,
both in the charged and neutral current sectors. The
Lorentz structure of the leptonic $\tau$ decays is certainly not determined,
but begins to be experimentally explored.  The quality of the hadronic 
$\tau$ decay data
has made possible to perform quantitative QCD tests
and determine the strong coupling constant very accurately.
Searches for non-standard phenomena have been pushed to the limits 
that the existing data samples allow to investigate.

At present, all experimental results on the $\tau$ lepton are consistent with
the Standard Model. There is, however, large room for improvements. 
Future $\tau$ experiments
will probe the Standard Model
to a much deeper level of sensitivity and will explore the
frontier of its possible extensions. 

\goodbreak
\nonumsection{ACKNOWLEDGEMENTS}

I am indebted to Andreas H\"ocker, Wolfgang Lohmann 
and Matthias Neubert for their help with the figures.
This work has been supported in part by CICYT (Spain) under grant 
No. AEN-96-1718.

\nonumsection{REFERENCES}\vspace*{-0.3cm}

\end{document}